\documentclass[10pt,journal,compsoc]{IEEEtran}
\ifCLASSOPTIONcompsoc
  \usepackage[nocompress]{cite}
\else
  \usepackage{cite}
\fi
\ifCLASSINFOpdf
\else
\fi
\usepackage{amsmath}
\usepackage{hyperref}
\urldef{\urlA}\url{https://github.com/nimish15shah/GRAPHOPT}

\usepackage[shortlabels]{enumitem}

\usepackage[ruled,vlined,linesnumbered]{algorithm2e}

\usepackage{amssymb}

\usepackage{amsmath}
\usepackage{graphicx}
\usepackage{multirow}
\usepackage{color, soul}
\usepackage{enumitem}
\newcommand{\ns}[1]{\textcolor{red}{[ns: #1]}}
\newcommand{\mv}[1]{{\textcolor{green}{[mv: #1]}}}
\newcommand{\wm}[1]{{\textcolor{cyan}{[wm: #1]}}}
\newcommand{\tool}{\textsc{\textsc{GraphOpt}}}

\newcommand{\rev}[1]{#1}
\newcommand{\rtwo}[1]{#1}

\soulregister{\ns}{1}
\soulregister{\mv}{1}
\soulregister{\wm}{1}

\newcommand{\Psymbol}{\textit{P}}

\usepackage{lipsum} 
\usepackage{dblfloatfix} 

\usepackage{listings, multicol}

\definecolor{ForestGreen}{RGB}{34,106,46}
\definecolor{lightgray}{rgb}{0.97, 0.97, 0.97}

\lstdefinelanguage{minizinc}{
    morekeywords={
        ann, annotation, any, array, assert,
        bool,
        constraint,
        else, elseif, endif, enum, exists,
        float, forall, function,
        if, in, include, int,
        list,
        minimize, maximize,
        of, op, output,
        par, predicate,
        record,
        set, solve, string,
        test, then, tuple, type,
        var,
        where,
        abort, abs, acosh, array_intersect, array_union,
        array1d, array2d, array3d, array4d, array5d, array6d, asin, assert, atan,
        bool2int,
        card, ceil, combinator, concat, cos, cosh,
        dom, dom_array, dom_size, dominance,
        exp,
        fix, floor,
        index_set, index_set_1of2, index_set_2of2, index_set_1of3, index_set_2of3, index_set_3of3,
        int2float, is_fixed,
        join,
        lb, lb_array, length, let, ln, log, log2, log10,
        min, max,
        pow, product,
        round,
        set2array, show, show_int, show_float, sin, sinh, sqrt, sum,
        tan, tanh, trace,
        ub, and ub_array,
        bool_search, int_search, seq_search, priority_search,
        minisearch, search, while, repeat, next, commit, print, post, sol, scope, time_limit, break, fail
    },
    sensitive=true, 
    morecomment=[l][\em\color{ForestGreen}]{\%},
    morestring=[b]",
}

\begin{document}
%
\title{\tool{}: constrained-optimization-based parallelization of irregular graphs}
%
%
%
%
 \author{Nimish~Shah, Wannes~Meert,~\IEEEmembership{Member,~IEEE},
         and~Marian~Verhelst,~\IEEEmembership{Senior Member,~IEEE}%
 \IEEEcompsocitemizethanks{\IEEEcompsocthanksitem N. Shah and M. Verhelst are with the Department
 of Electrical Engineering - MICAS, KU Leuven, Belgium.\protect\\
 \IEEEcompsocthanksitem W. Meert is with the Department
 of Computer Science - DTAI, KU Leuven, Belgium.}
}
\IEEEtitleabstractindextext{%
\begin{abstract}
 Sparse, irregular graphs show up in various applications like linear algebra, machine learning, engineering simulations, robotic control, etc. These graphs have a high degree of parallelism, but their execution on parallel threads of modern platforms remains challenging due to the irregular data dependencies. The execution performance can be improved by efficiently partitioning the graphs such that the communication and thread synchronization overheads are minimized without hurting the utilization of the threads. To achieve this, this paper proposes \tool{}, a tool that models the graph parallelization as a constrained optimization problem and uses the open Google OR-Tools solver to find good partitions. Several scalability techniques are developed to handle large real-world graphs with millions of nodes and edges. Extensive experiments are performed on the graphs of sparse matrix triangular solves (linear algebra) and sum-product networks (machine learning), respectively, showing a  mean speedup of 2.0$\times$ and 1.8$\times$ over previous state-of-the-art libraries, demonstrating the effectiveness of the constrained-optimization-based graph parallelization.
\end{abstract}

}

\maketitle

\IEEEdisplaynontitleabstractindextext

%
\IEEEpeerreviewmaketitle

\IEEEraisesectionheading{\section{Introduction}\label{sec:intro}}

%
%
%
%
\IEEEPARstart{S}{ignificant} advances have been made in accelerating massively parallel computations like image processing and deep neural networks, which typically involve \emph{regular} memory accesses
and well-structured explicit parallelism.
These advances can be ascribed to dedicated programming frameworks like Tensorflow \cite{45381} and Halide \cite{DBLP:conf/pldi/Ragan-KelleyBAPDA13}, optimized compilation approaches like polyhedral compilation \cite{DBLP:conf/sc/PradelleMBSL17}, and specialized hardware architectures like GPU and TPU. However, these approaches cannot be readily extended to another equally important class of parallel applications that is \emph{irregular}. Such applications involve irregular memory accesses lacking any apparent recurring patterns, and parallelism that is difficult to partition. Such computations arise in applications like graph workloads, sparse linear algebra, probabilistic machine learning, recommender systems etc.

\begin{figure}[!t]
\centering
\includegraphics[trim={0cm 0.5cm 0cm 0cm} , clip, width=0.8\columnwidth]{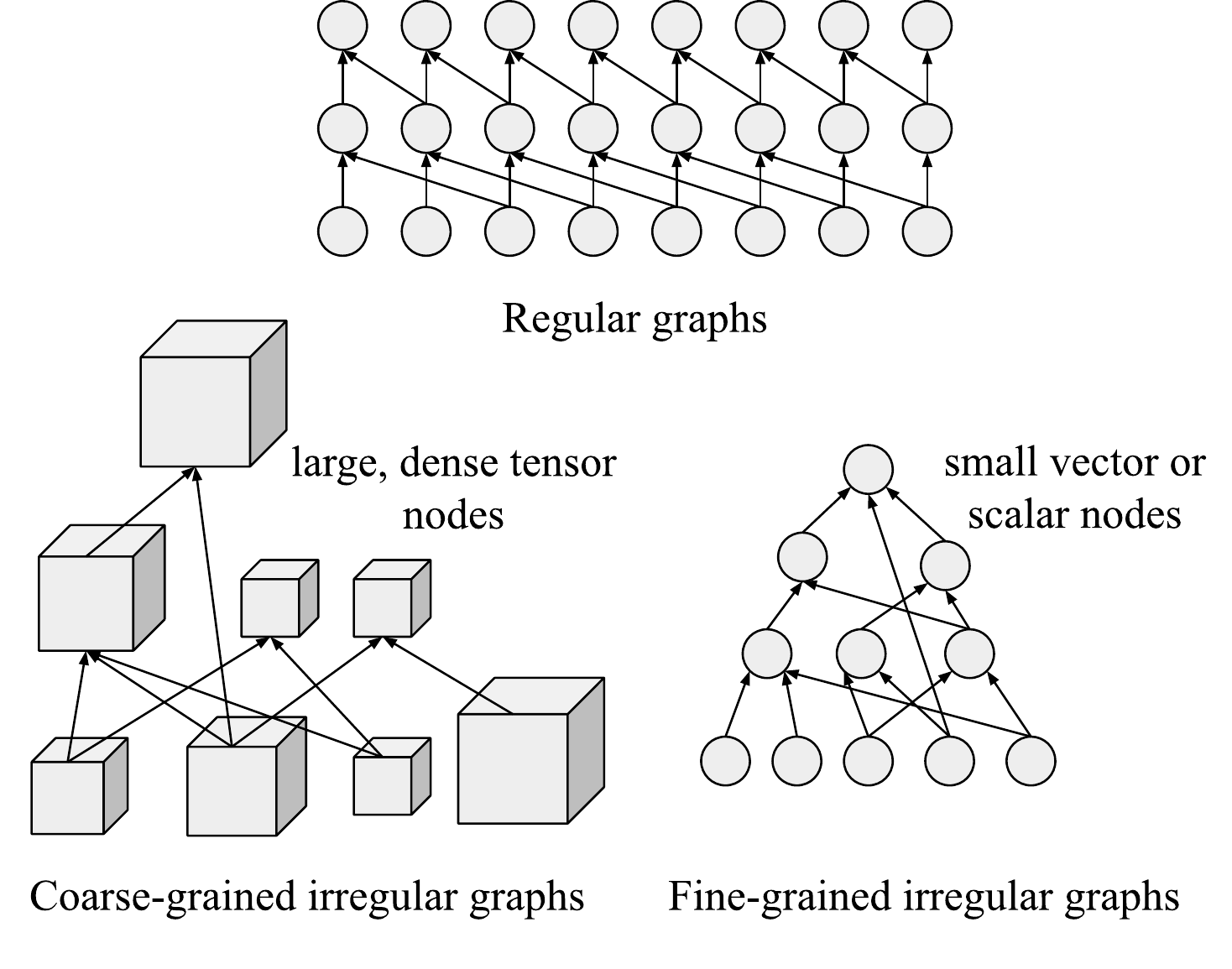}
\caption{\textbf{DAG structures and granularities}. Illustration of regular, coarse-grained irregular, and fine-grained irregular compute graphs}%
\label{fig:coarse_vs_fine}
\end{figure}

A subset of these irregular applications can be represented as static directed acyclic graphs (DAG), where the nodes represent operations to perform and edges represent data dependencies among these operations. These DAG workloads can be further classified into two categories: 1) coarse-grained and 2) fine-grained, as shown in fig. \ref{fig:coarse_vs_fine}. The coarse-grained DAGs, although irregular, have as their nodes large regular operations like dense tensor operations. The computation within each node is enough to amortize parallelization costs associated with thread synchronization, inter-thread communication, GPU kernel launch, etc. The examples of coarse-grained graphs include neural networks and linear algebra with block/structured sparsity,
graph neural networks, visual rendering, electronic circuit placement, video encoding etc. These applications can be accelerated efficiently using frameworks like Tensorflow, TaskFlow \cite{DBLP:conf/ipps/HuangLGW19}, Intel Thread Building Blocks (TBB) \cite{Robison2011} etc.,
by modeling nodes as individual tasks.

On the other hand, 
nodes in \textit{fine-grained} DAGs represent only a couple of scalar operations, whose computation cost cannot amortize synchronization and task launch overheads when each node is modeled as an individual task.
As a result, these DAGs cannot be accelerated by simply modeling them as task graphs in Tensorflow, etc. Their acceleration needs the creation of coarser partitions by combining original fine-grained nodes, to increase computation to synchronization/communication ratio. But if the partitions become \textit{too} coarse, they hurt parallelism as there might not be enough partitions available to execute in parallel. Hence, appropriate granularity and parallelism of the partitions are critical for good acceleration.
To this end, this paper proposes \textbf{\tool{}}\footnote{Available at \urlA
}---a partitioner to efficiently parallelize fine-grained DAGs through hardware-aware partitioning. The key contributions of the paper are as follows: 
\begin{itemize}
    \item \tool{} models the graph partitioning for parallel execution as a constrained optimization problem, and leverages a state-of-the-art solver. 
    \item Several scalability techniques are proposed to handle real-world graphs with millions of nodes and edges.
    \item The performance of \tool{} is validated for two applications: 1) sparse matrix triangular solves from linear algebra and 2) sum-product networks from machine learning, and compared against standard libraries.
\end{itemize}
The paper is organized as follows. Section \S \ref{sec:problem} defines the problem of parallelizing fine-grained DAGs and \tool{}'s approach. Section \S \ref{sec:graph_opt} details the working of \tool{}, followed with extensive performance benchmarking in section \S \ref{sec:experiments}. Finally, section \S \ref{sec:related_works} discusses related works and section \S \ref{sec:conclusion} concludes the paper.

\begin{figure}[!t]
\centering
\includegraphics[trim={0cm 0cm 0cm 0cm} , clip, width=0.95\columnwidth]{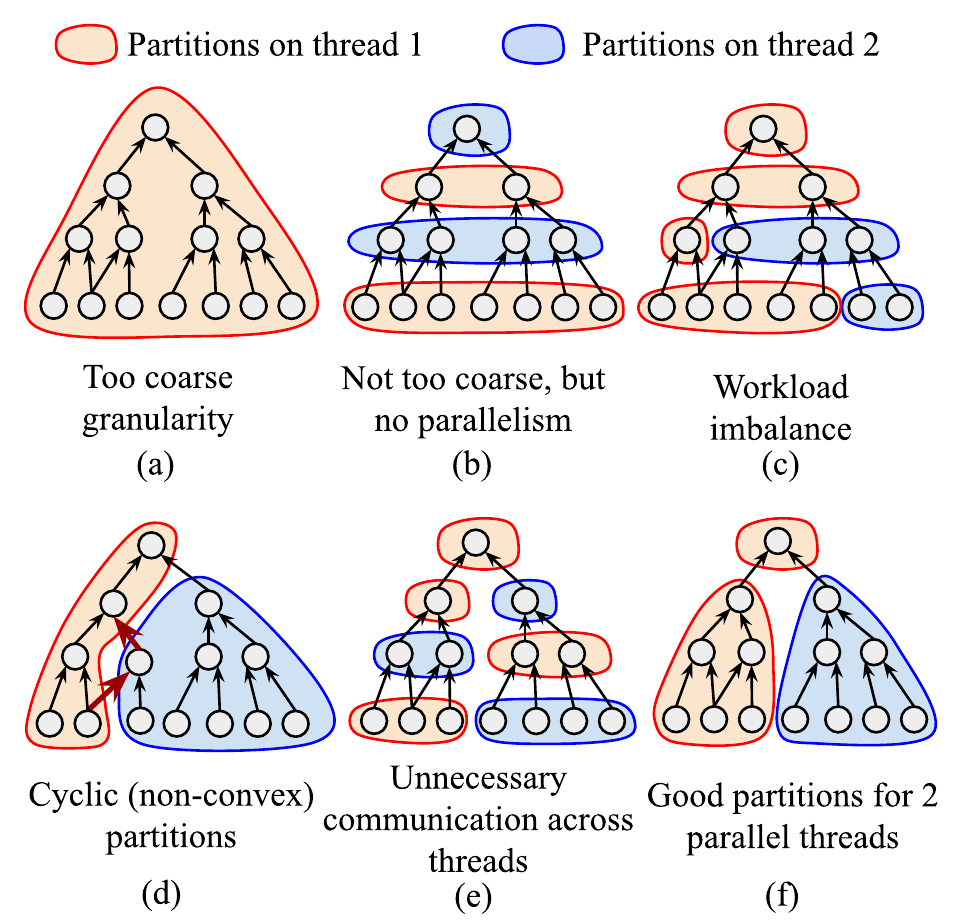}
\caption{\textbf{Different ways to partition a DAG}, and impact on parallelization, communication, and workload balance}%
\label{fig:granularity_examples}
\end{figure}%

\section{Graph partitioning for parallelization} \label{sec:problem}
\begin{figure}[!t]
\centering
\includegraphics[trim={0cm 0cm 0cm 0cm} , clip, width=0.95\columnwidth]{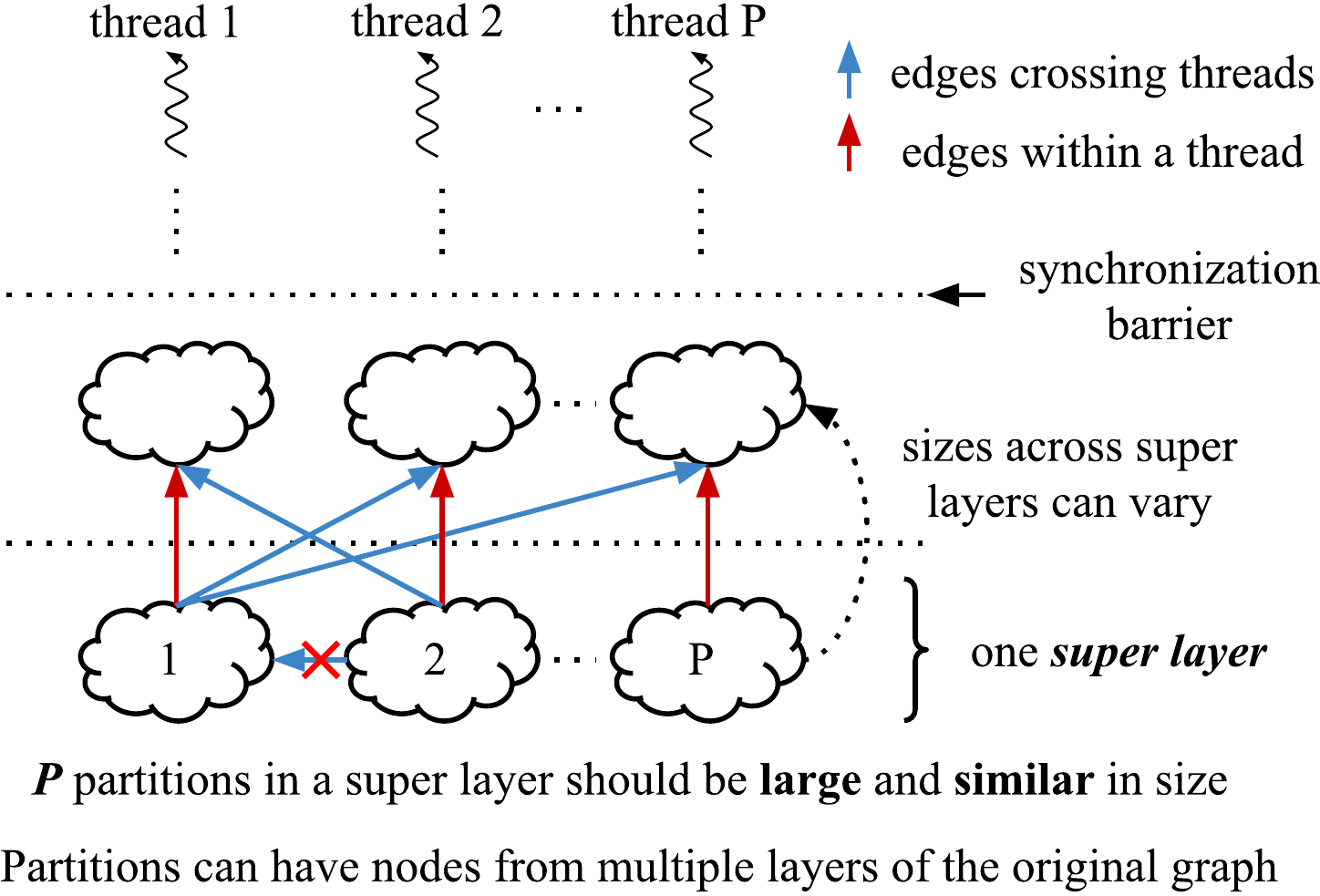}
\caption{\textbf{Super layers}. \tool{} decomposes a fine-grained DAG into \textit{super layers}, each having \Psymbol{} partitions. The partitions are made as large as possible but also of similar size to ensure workload balancing}%
\label{fig:super_layer}
\end{figure}

Modern computing platforms like multi-core CPUs and GPUs are equipped with parallel hardware threads, which can execute DAG nodes in parallel. However, the nodes cannot be arbitrarily scheduled because of the data dependencies represented by the DAG edges, i.e., a node can only be executed after all the \textit{predecessor} nodes have finished their execution. This data dependency demands threads synchronization when a predecessor node is mapped to a different thread than the current one. Furthermore, the nodes in raw fine-grained DAGs do not have enough computation to be executed as a stand-alone task in a task-scheduling framework (like Intel TBB \cite{Robison2011}, etc.)
because of the task-launch and synchronization overheads. This can be addressed by clustering the nodes in \emph{coarser partitions} that are executed as monolithic tasks to amortize the overheads. The quality and correctness of these coarser partitions depend on several aspects as discussed next (with illustrations in fig. \ref{fig:granularity_examples}).
\newline\\
\textbf{Data dependencies and acyclic partitions}: The data dependencies of DAG edges imply that
a partition can be launched as a monolithic unit only if all the predecessor partitions have finished the execution. Furthermore, the edges between partitions should not create cycles, because cyclic partitions need intermittent synchronization to resolve intermediate data dependencies. The cycles can be prevented by generating \emph{convex} partitions, i.e., if two nodes are in a partition, the nodes on all the (directed) paths in between the two nodes also have to be in the partition. An example of cyclic partitions is shown in the fig.~\ref{fig:granularity_examples}(d).
\tool{} avoids these cycles by enforcing an acyclic constraint. 
\newline\\
\textbf{Granularity and parallelism}: The sizes of the partitions dictate how often the threads need to synchronize and communicate. To reduce the overall runtime, there are multiple, conflicting, requirements:%
\begin{itemize}
    \item Partitions should be as large as possible to amortize the overheads and increase the local data reuse, minimizing communication across threads. Eg., increasing the granularity in fig.~\ref{fig:granularity_examples}(c) and (e) helps.
    \item There should be enough parallel partitions to keep all the available threads in the underlying hardware busy. Partitions in fig.~\ref{fig:granularity_examples}(a) and (b) do not have any parallelism across threads at all.
    \item Partitions that are supposed to execute in parallel should be of similar size to balance the workload across threads. Figure \ref{fig:granularity_examples}(c) shows a counter-example.
\end{itemize}%
These requirements suggest that the partition granularity depends on the parallelism of the underlying hardware like the number of CPU threads, which is fixed and known beforehand. It also depends on the available parallelism in the DAG, which can vary in the different parts of the DAG. Hence, the partition granularity cannot be manually fixed but has to be automatically adjusted depending on the DAG structure. \tool{} achieves a balance by generating partitions that are as large as possible, as long as there are enough parallel partitions.
\newline\\
\textbf{Communication}: An edge from one partition to the other indicates communication of data. If the two partitions are executed on the same thread,
the data can be locally reused via local caches or scratchpads. Whereas, the communication across threads
incurs a higher overhead of flushing data to the shared caches. This communication overhead can be reduced by keeping the edges within the same thread as much as possible. Figure \ref{fig:granularity_examples}(e) shows an example where partitions can be simply remapped to different threads to avoid inter-thread communication.
\newline\\
\textbf{\tool{}'s approach}: To appropriately handle the various trade-offs, instead of creating arbitrary partitions, \tool{} creates a layered graph of coarse partitions as shown in fig. \ref{fig:super_layer}. To avoid confusion, a layer of coarse partitions is always referred to as a \textit{\textbf{super layer}} in the rest of the paper, and a \textit{\textbf{layer}} simply means a layer of nodes in the original fine-grained DAG. Every super layer has \Psymbol{} independent partitions, where the user can specify an arbitrary \Psymbol{} based on the target hardware (e.g. \Psymbol{} can be set to 8 for a CPU with 8 single-threaded cores). This ensures that, whenever possible, all the \Psymbol{} hardware threads have a corresponding partition to execute. The partitions in a super layer are independent, i.e. there is no edge crossing among them, enabling parallel execution. The parallel threads need to be synchronized after every super layer to allow race-free data communication required by the blue edges. \tool{} models this multi-objective problem of DAG partitioning to create super layers as a constrained optimization problem, as explained in the next section.
\section{\tool{}} \label{sec:graph_opt}
This section explains the generation of super layers with \Psymbol{} parallel, balanced, coarse partitions from a fine-grained DAG. The optimal graph partitioning, even without the acyclic constraint and parallelism requirement, is known to be NP-complete \cite{DBLP:journals/tcs/Feldmann13, DBLP:journals/siamsc/KarypisK98}. Recent work \cite{DBLP:journals/heuristics/MoreiraPS20} has shown that graph partitioning with the acyclic constraint is also NP-complete. Section~\S \ref{sec:related_works} explains why general graph partitioning approaches cannot be used here because of the parallelism requirement (\Psymbol{} independent partitions) and the acyclic constraint. These general approaches only focus on minimizing edges among (possibly cyclic) partitions, and do not aim to generate layered graphs with P parallel workload-balanced partitions needed to keep parallel threads busy.

\begin{figure}[!t]
\centering
\includegraphics[trim={0cm 0cm 0cm 0cm} , clip, width=\columnwidth]{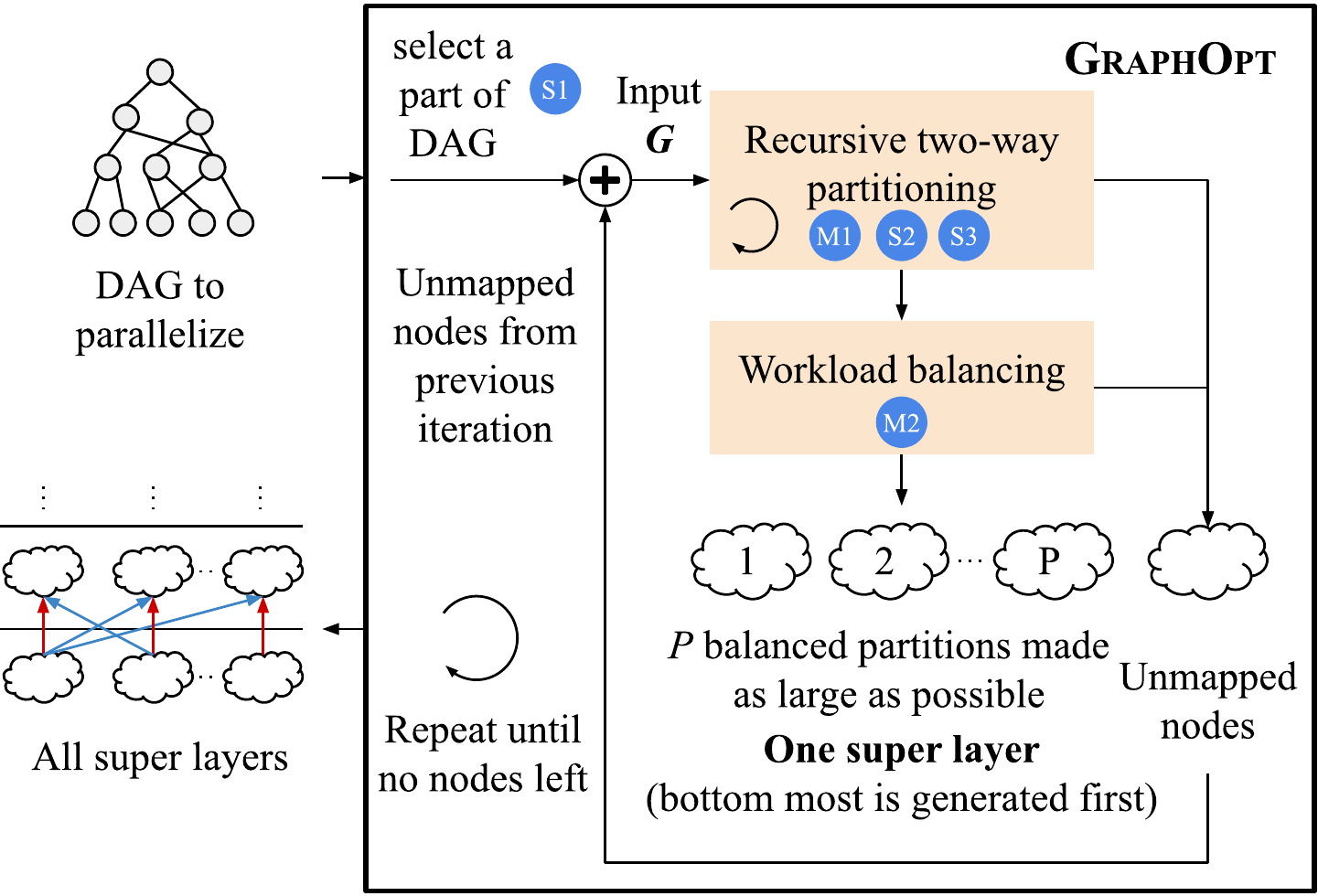}
\caption{\textbf{\tool{} overview}. One super layer, starting from the bottom, is generated in every iteration. The main steps M1 and M2 use an optimization model with the Google OR-tool solver \cite{ortools} to find good partitions. The scalability steps S1, S2, and S3 are used to handle graphs with millions of nodes/edges.}%
\label{fig:flow_summary}
\end{figure}

Figure \ref{fig:flow_summary} shows \tool{}'s approach for generating partitioned super layers. To reduce the complexity, instead of finding all the super layers simultaneously, the tool iteratively constructs 
one super layer at a time going from the bottom to the top. In an iteration, a super layer is generated with two main steps: (M1) recursive two-way partitioning 
and (M2) workload-balancing
. The S1, S2, and S3 are the 
scalability techniques that enable \tool{} to handle large graphs with millions of nodes/edges. \footnote{The full algorithmic description is available in appendix \ref{sec:appendix_a}}

An iteration starts by selecting a part of the DAG that is considered for generating the current super layer. Ideally, all the currently unmapped nodes should be considered, but to limit the complexity, step S1 selects a subgraph \emph{G} from the unmapped DAG. The M1 step splits this subgraph \emph{G} into \Psymbol{} parallel partitions and associates them to the \Psymbol{} underlying hardware threads. These partitions, however, could potentially be of different sizes. The M2 step redistributes nodes among these partitions for workload balancing, generating \Psymbol{} balanced partitions for the current super layer. The unmapped nodes that could not be mapped to any partitions are then considered for the subsequent super layers.


\begin{figure}[!t]
\centering
\includegraphics[trim={0cm 0cm 0cm 0cm} , clip, width=0.75\columnwidth]{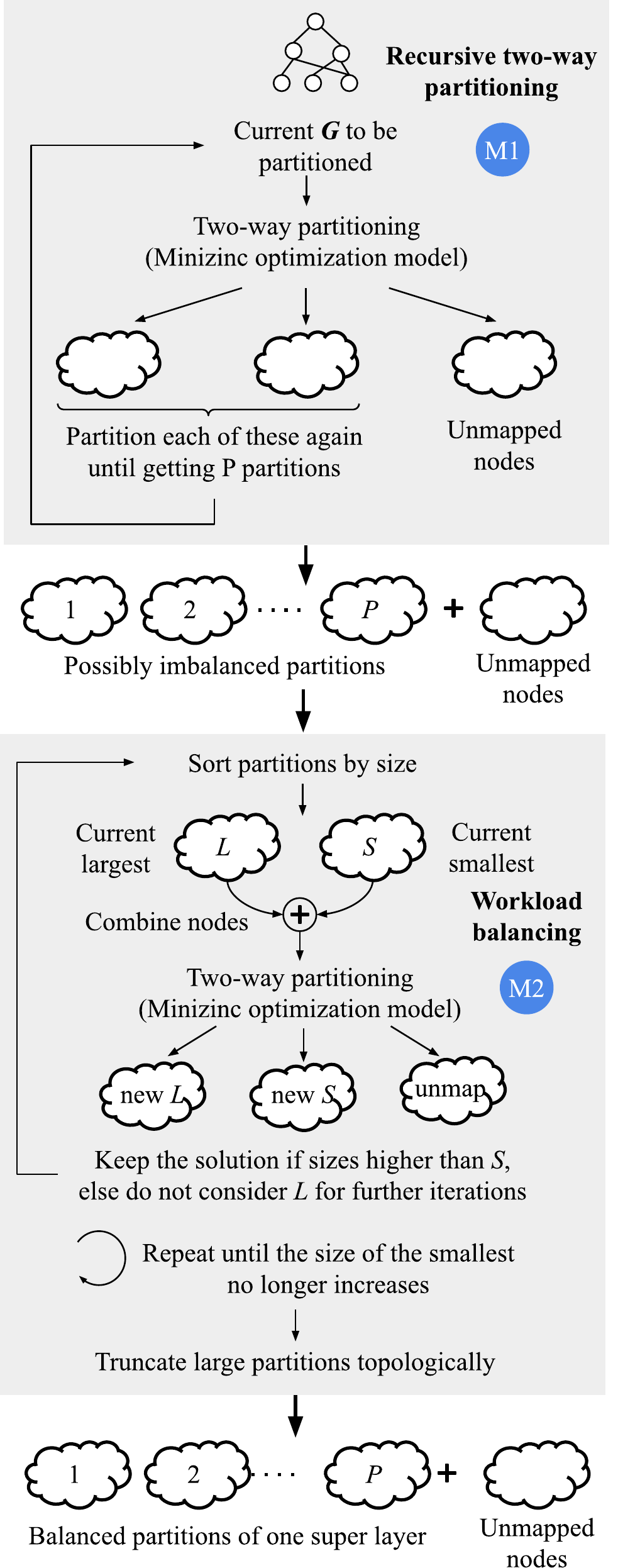}
\caption{The \textbf{recursive two-way partitioning (M1)} uses a Minizinc-based optimization model to partition subgraphs until getting \Psymbol{} partitions. Due to the recursive approach, the partitions can be imbalanced. The \textbf{workload balancing (M2)} iteratively redistributes the nodes, using the same Minizinc-based model, to generate a balanced super layer}%
\label{fig:flow_M1_M2}
\end{figure}
\subsection{Recursive two-way partitioning (M1)} \label{sec:M1}
Given a subgraph \emph{G}, a super layer with parallel partitions should ideally be constructed by a direct \textit{P-way} partitioning. \tool{} relaxes this by using recursive \textit{two-way} partitioning, i.e. recursively splitting subgraphs into two parallel smaller subgraphs until getting \Psymbol{} subgraphs (partitions) for \Psymbol{} threads, as illustrated in fig.~\ref{fig:flow_M1_M2}(top). The recursion starts with the aim of mapping the input subgraph \emph{G} to \Psymbol{} threads. The first two-way partition generates two output partitions, one of which represents the nodes corresponding to {$t_1,...,t_{P/2}$} threads, and the other corresponding to {$t_{P/2+1},...,t_{P}$} threads. The third output is a set of unmapped nodes that cannot be mapped to either partition without adding an edge between the partitions. These unmapped nodes will be considered for the subsequent super layers. The next recursion splits the first partition (which becomes the current \emph{G}) into two smaller partitions, one for {$t_{1},...,t_{P/4}$} threads and the other for {$t_{P/4+1},...,t_{P/2}$} threads. This repeats until partitions for individual threads are determined. Thus the problem of one super layer generation is reduced to multiple iterations of two-way partitioning under given constraints and objectives. 
Note that if \Psymbol{} is not a power of 2, the partitions can end up being imbalanced, which is addressed with the M2 step.

Instead of developing a custom heuristic algorithm, \tool{} models the two-way partitioning problem with the Minizinc constrained-optimization language \cite{DBLP:conf/cp/NethercoteSBBDT07}.
Such a model can be solved with state-of-the-art constraint programming (CP) or mixed-integer linear programming (MILP) solvers like Google OR-Tools \cite{ortools}, SCIP \cite{GamrathEtal2020OO}, Gurobi \cite{gurobi} etc. These solvers are designed and tuned with decades of research and often outperform custom heuristics for smaller problem instances, but struggle to scale to larger problems. This is addressed in this work by developing the scalability techniques explained later in sec. \ref{sec:S1_S2_S3}. An optimization model consists of four parts: 1) inputs of the problem, 2) decision variables whose values need to be determined by the solver, 3) constraints on the decision variables, and 4) the objective of the optimization (refer \cite{DBLP:series/synthesis/2013Sankaralingam} for more details of constrained optimization). The rest of the section describes these four parts.
\newline\\
\textbf{3.1.1 Optimization model for the two-way partitioning}\\
Given an input graph, the two-way partitioning aims to allocate the nodes to two parallel partitions such that there is no edge crossing from one to the other. Furthermore, the size of the partitions should be as large and equal as possible. If the target for the current recursion are {$t_1, t_2,...,t_x$} threads, the first output partition corresponds to {$t_1,..., t_{x/2}$} threads and other to {$t_{x/2+1},..., t_x$} threads. The inter-thread communication (blue edges in fig. \ref{fig:super_layer}) would reduce if the sources of the incoming edges of the partitions are from the same group of threads. Hence, one of the aims of the two-way partitioning is also to perform allocation such that the nodes in the first partition mostly have incoming edges from {$t_1,..., t_{x/2}$} threads, and the second partition from {$t_{x/2+1},..., t_x$} threads.

Table \ref{tab:model} shows different pasts of the optimization model. The current input DAG is denoted by \emph{G}(\emph{V}, \emph{E}), where \emph{V} is the set of nodes (vertices), and \textit{E} is the set of directed edges. Note that this is not the complete original DAG, but an input subgraph for the current recursion of the two-way partitioning. The output partitions of \emph{G} are decided by the decision variable \emph{PART}(\emph{V}), an array of integers, one for every node. If \emph{PART}[\emph{v}]= 1 or 2, the node \emph{v} is allocated to the partition 1 or 2 respectively, and if \emph{PART}[\emph{v}]= 0, it is not allocated. 
The constraints and objectives are discussed next.
\rev{\textbf{Notations}: $\forall$ for all, $\in$  is member of, $|$ such that, $\lor$ logical \textit{or}, $\land$ logical \textit{and}, $\lfloor \; \rfloor$ floor function.}
\\~\\
\textbf{Acyclic and data-dependency constraint}\\
There should not be any edge from one partition to the other, which implies that, for the destination and source nodes of every edge, \emph{PART}[\emph{dst}] should either be equal to \emph{PART}[\emph{src}] or should be 0. This also ensures that the destination node is unallocated if the source node is also unallocated, which is needed because the edges represent data-dependencies. This constraint is modeled as follows,
\renewcommand{\arraystretch}{1.3}
\begin{table*}
\centering
\caption{\rev{\textbf{Optimization model}. Inputs, decision variables, constraints, and objective of the Minizinc two-way partitioning optimization model}}%
\begin{tabular}{|l|l|l|}
\hline
\multicolumn{1}{|c|}{\textbf{Name}} & \multicolumn{1}{c|}{\textbf{Description}} & \multicolumn{1}{c|}{\textbf{Type}} \\ \hline
\multicolumn{3}{|c|}{\textbf{Inputs}} \\ \hline
$t_1$, ..., $t_x$ & Target threads for this recursion & array of int \\ \hline
\emph{V} & Set of nodes in the current DAG \emph{G} & set of int \\ \hline
\emph{E} (\emph{V}, \emph{V}) & Set of directed edges in \emph{G} expressed as tuples of source and destination nodes & set of int tuples \\ \hline
\emph{node\_w} (\emph{V}) & Weights indicating the amount of computation within nodes & array of int \\ \hline
$V_{in}$ & Set of nodes that are already allocated to previous super layers & set of int \\ \hline
$E_{in}$ ($V_{in}$, \emph{V}) & Set of directed edges with source node in $V_{in}$ and destination in \emph{V} & set of int tuples \\ \hline
$\mathit{PART_{in}}$ (\emph{$V_{in}$}) & Allocated partitions of nodes in \emph{$V_{in}$} & array of int in {[}1,2{]} \\ \hline
\multicolumn{3}{|c|}{\textbf{Decision vairables: Final output}} \\ \hline
\emph{PART} (\emph{V}) & Allocated partitions for nodes in \emph{V}, where 0 indicates no allocation & array of int in {[}0,2{]} \\ \hline
\multicolumn{3}{|c|}{\textbf{Decision variables: Intermediate}} \\ \hline
\emph{PART\_1\_size} & Amount of computation allocated to partition 1 & int \\ \hline
\emph{PART\_2\_size} & Amount of computation allocated to partition 2 & int \\ \hline
$\mathit{E_{in}\_crossing}$ ($E_{in}$) & Indicates whether the edge in $E_{in}$  are crossing partitions or not & array of bool \\
\hline
\multicolumn{3}{|c|}{\textbf{Constraints}} \\ \hline

Acyclic parts & \multicolumn{2}{|c|}{ $\forall{(src, dst)} \in E$,
$\mathit{PART}[dst] = \mathit{PART}[src] \; \lor \; \mathit{PART}[dst] = 0$ } \\ 
\hline

Partition size & 
\multicolumn{2}{|c|}{
$\forall{v} \in V, \; \mathit{PART\_1\_size} = \textrm{sum}(node\_w[v] \;\vert\; \textrm{ if } \mathit{PART}[v] = 1)$,
}
\\

& 
\multicolumn{2}{|c|}{
$\; \; \; \; \; \mathit{PART\_2\_size} = \textrm{sum}(node\_w[v] \;\vert\; \textrm{ if } \mathit{PART}[v] = 2)$ 
} \\ 
\hline

Inter-thread comm.&
\multicolumn{2}{|c|}{
$\forall{(src, dst)} = e \in E_{in}, \;
\mathit{E_{in}\_crossing}[e] =  ((\mathit{PART}[dst] \neq 0) \; \land \; (\mathit{PART}[dst] \neq \mathit{PART_{in}}[src])) \label{eq:edge_crossing_constraint}
$
}
\\
\hline

\multicolumn{3}{|c|}{\textbf{Objective}} \\ \hline

\multicolumn{3}{|c|}{ $\textrm{maximize \;\;} w_s \times \textrm{min}(\mathit{PART\_1\_size}, \;\mathit{PART\_2\_size})
\; - \; w_c \times \textrm{sum}(E_{in}\_crossing)$ }\\ \hline
\end{tabular}%
\label{tab:model}%
\end{table*}%
\renewcommand{\arraystretch}{1.0}
{
\begin{gather}
\forall{(src, dst)} \in E, \nonumber\\
\mathit{PART}[dst] = \mathit{PART}[src] \; \lor \; \mathit{PART}[dst] = 0
\label{eq:acyclic_constraint}
\end{gather}
}%
\textbf{Partition size constraint}\\
The input parameter $\mathit{node\_w}$($V$) (array of integers) represents the amount of computation to be performed in each node.  For workload balancing, the amount of computation in the two partitions should be the same. To formulate such an optimization objective, the decision variables $\mathit{PART\_1\_size}$ and $\mathit{PART\_2\_size}$ are used, which are modeled with the following constraint,
{
\begin{gather}
\forall{v} \in V, \nonumber\\
\mathit{PART\_1\_size} = \textrm{sum}(node\_w[v] \;\vert\; \textrm{ if } \mathit{PART}[v] = 1) \nonumber\\
\mathit{PART\_2\_size} = \textrm{sum}(node\_w[v] \;\vert\; \textrm{ if } \mathit{PART}[v] = 2) 
\end{gather}
}%
\textbf{Inter-thread communication constraint}\\
The blue edges in fig. \ref{fig:super_layer} contribute to inter-thread communication. An edge would become a blue edge if its source node is already allocated in a previous super layer to a thread $t_a$, and the destination node is currently allocated to a partition that does not correspond to $t_a$. Note that these edges are not in the current edge set $E$, because the edges in $E$ have the source and destination nodes that are being considered for the current super layer. Another input to the model $E_{in}$ represents such edges, which have source nodes (represented by the set $V_{in}$) in the previously generated super layers, and destination nodes in the current $G$. 
Input $\mathit{PART_{in}}$($V$) represents the partitions of the nodes in $V_{in}$ based on the threads they are allocated to. If a $v_{in}$ was allocated to any of the threads {$t_1,..., t_{x/2}$} in the previous super layers, $\mathit{PART_{in}}$[$v_{in}$] = 1, and if allocated to any of the {$t_{x/2+1},..., t_x$} threads $\mathit{PART_{in}}$[$v_{in}$] = 2. If a $v_{in}$ was not allocated to any of the current target threads {$t_1,..., t_x$}, its corresponding edges always result in inter-thread communication irrespective of how the current partitioning is done.
Hence, such $v_{in}$ and the corresponding edges are not considered in the model.  

\begin{figure}[!t]
\centering
\includegraphics[trim={0cm 0cm 0cm 0cm} , clip, width=\columnwidth]{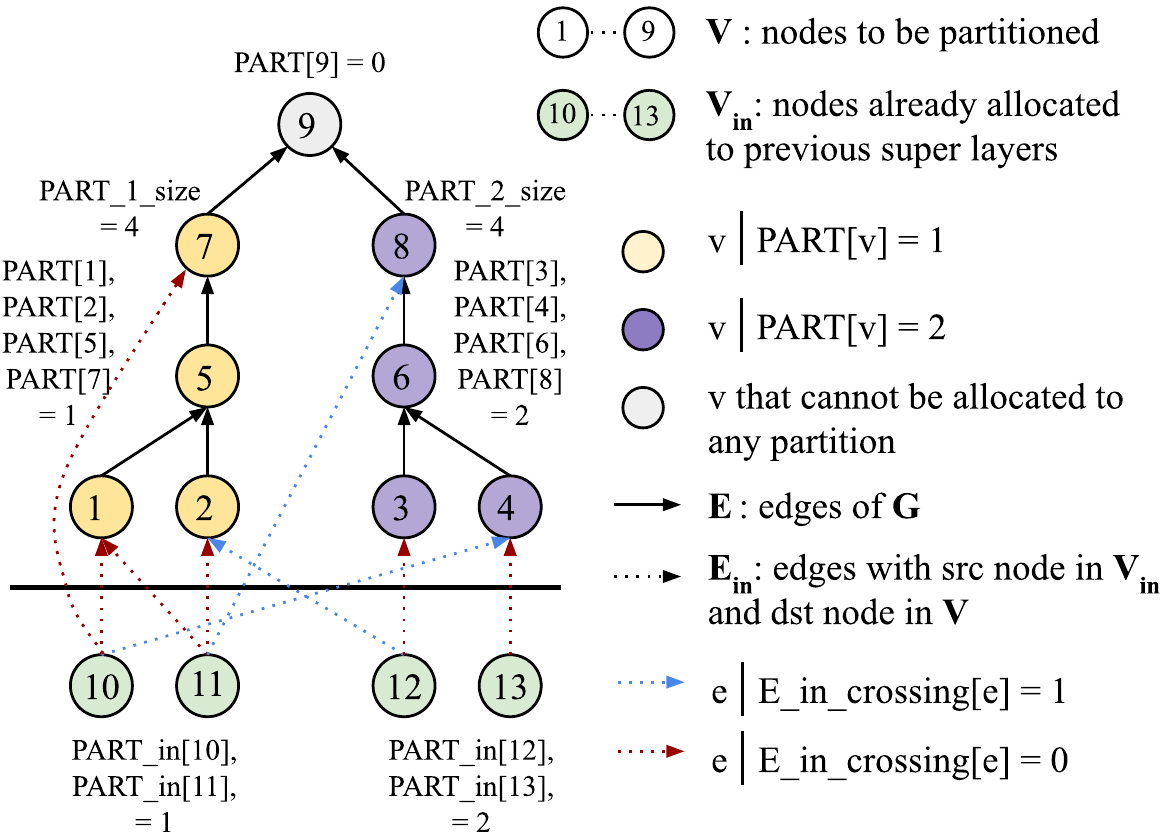}
\caption{\textbf{Example}. An optimal partitioning for a simple graph}%
\label{fig:example}
\end{figure}

An edge in $E_{in}$ is a \emph{crossing} edge when the partions of the destination and source nodes are different. The edge crossings are tracked with $\mathit{E_{in}\_crossing}$ (array of booleans), modeled with the following constraint,
{
\begin{gather}
\forall{(src, dst)} = e \in E_{in}, \;
\mathit{E_{in}\_crossing}[e] = \nonumber\\ ((\mathit{PART}[dst] \neq 0) \; \land \; (\mathit{PART}[dst] \neq \mathit{PART_{in}}[src])) \label{eq:edge_crossing_constraint}
\end{gather}
}%
The first inequality makes sure that the edges with unallocated destination are not marked as crossing edges.
\newline \\
\textbf{Objective}\\
The objectives of \tool{} are maximizing but also equalizing the partition sizes and minimizing the edge crossings. 
The optimization solvers generally support only one objective, hence a weighted sum of the multiple objectives is used. The objective makes sure the smaller of the two partitions is made as big as possible. 
{
\begin{align}
\textrm{maximize \;\;} & w_s \times \textrm{min}(\mathit{PART\_1\_size}, \;\mathit{PART\_2\_size})\nonumber\\
\; - \; & w_c \times \textrm{sum}(E_{in}\_crossing) \label{eq:objective}
\end{align}
}%
\rev{Typically, global synchronization cost is significantly higher than communication cost (i.e., accessing data from other core's cache), the hyperparameter $w_s$ should be set higher than $w_c$. In our experiments, $w_s$ is set to 10 $\times$ $w_c$.} The objective along with the constraints in the equations \ref{eq:acyclic_constraint}-\ref{eq:objective} form the Minizinc optimization model, 
\footnote{The optimization model in the Minizinc language is available in appendix \ref{sec:minizinc_code}}, 
which is solved with the Google OR-Tools solver.
\newline \\
\textbf{3.1.2 Example}\\
The figure \ref{fig:example} illustrates the optimal two-way partitioning of a simple $G$. The nodes set $V$ is \{1, ..., 9\} and the edge set $E$ is \{(1,5), (2,5), ..., (8,9)\}. The computation in every node, the $\mathit{node\_w}$, is assumed to be 1. Assume that the target threads are \{$t_1$, ..., $t_4$\}.
The incoming edges $E_{in}$ are \{(10,1), (10,7), ..., (13,4)\} and suppose their source nodes $V_{in}$, \{10, ..., 13\}, are mapped to the threads \{$t_2$, $t_2$, $t_4$, $t_3$\} respectively, in the previous super layers. Hence, $\mathit{PART_{in}}$[10] = $\mathit{PART_{in}}$[11] = 1, and $\mathit{PART_{in}}$[12] = $\mathit{PART_{in}}$[13] = 2. 

The variable $\mathit{PART}$[9] for the top node will remain 0,
otherwise due to the constraint in the eq. \ref{eq:acyclic_constraint}, all the nodes would end up in the same partition (which would be a valid but suboptimal solution). The $\mathit{PART}$ for the other nodes are decided such that the $\mathit{E_{in}\_crossing}$ (blue edges) are minimized. Switching the $\mathit{PART}$ of 1,2,5,7 with 3,4,6,8 would lead to more blue edges. Hence, the solution shown in the figure is the optimal solution. In one of the next recursions, $G$ will contain \{1,2,5,7\} nodes with the target threads \{$t_1$, $t_2$\}. And the other recursion will be with \{3,4,6,8\} nodes and target threads \{$t_3$, $t_4$\}. Node 9 is an unmapped node, which will return to the set of remaining nodes to be used in the construction of the next super layer.
\begin{figure*}[!t]
\centering
\includegraphics[trim={0cm 0cm 0cm 0cm} , clip, width=0.83\textwidth]{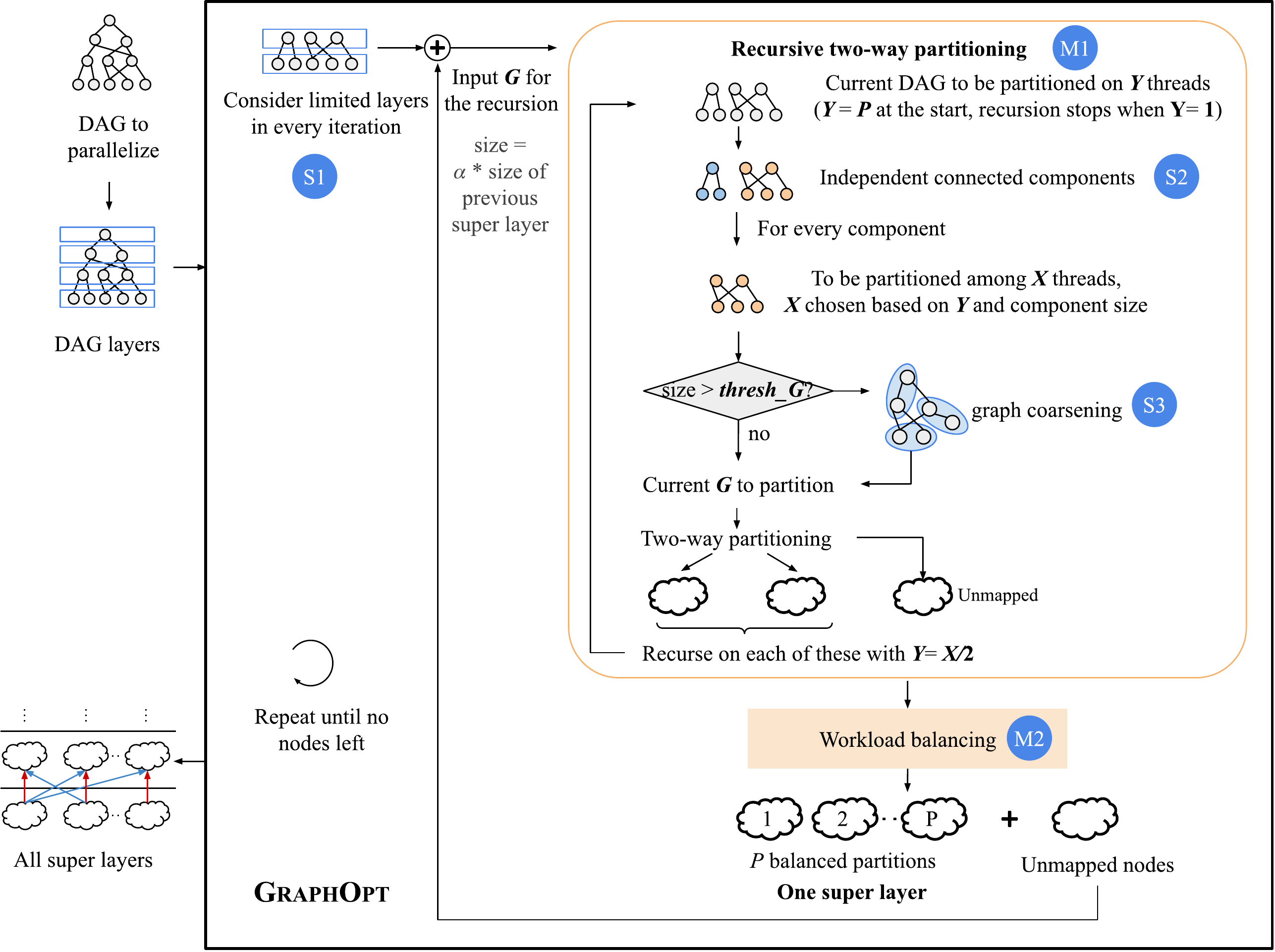}
\caption{\textbf{The full flow} with S1, S2, and S3 scalability steps that enables the \tool{} to handle large graphs with millions of nodes/edges.}%
\label{fig:flow_detailed}
\end{figure*}%
\subsection{Workload balancing (M2)} \label{sec:M2}
A penalty of using recursive two-way partitioning instead of a direct \Psymbol{}-way partitioning is that the \Psymbol{} partitions are not guaranteed to have the same size. The two-way partitioning attempts to equalize the partition sizes, but this does not guarantee equal parallelism. This imbalance can lead to divergences in partition sizes in subsequent recursions. Moreover, the imbalance can also occur when \Psymbol{} is not a power of 2. To address this, a workload balancing step (M2) is used as shown in fig.~\ref{fig:flow_M1_M2}(bottom), which redistributes the nodes across imbalanced partitions. 
In every iteration, the partitions are sorted according to the sizes. The nodes of the largest and the smallest partitions are combined and two-way repartitioned (with the Minizinc model) in an attempt to redistribute the workload. This repeats until the size of the smallest no longer increases further. If the sizes are still unequal, the larger partitions are truncated in topological order to equalize the sizes (with some margin). The truncated nodes will be added to the pool of unmapped nodes to be considered for the next super layer. The final output of this step is a super layer with balanced \Psymbol{} partitions.

\subsection{Scale to large graphs (S1, S2, S3)} \label{sec:S1_S2_S3}
\begin{figure}[!t]
\centering
\includegraphics[trim={0cm 0cm 0cm 0cm} , clip, width=0.7\columnwidth]{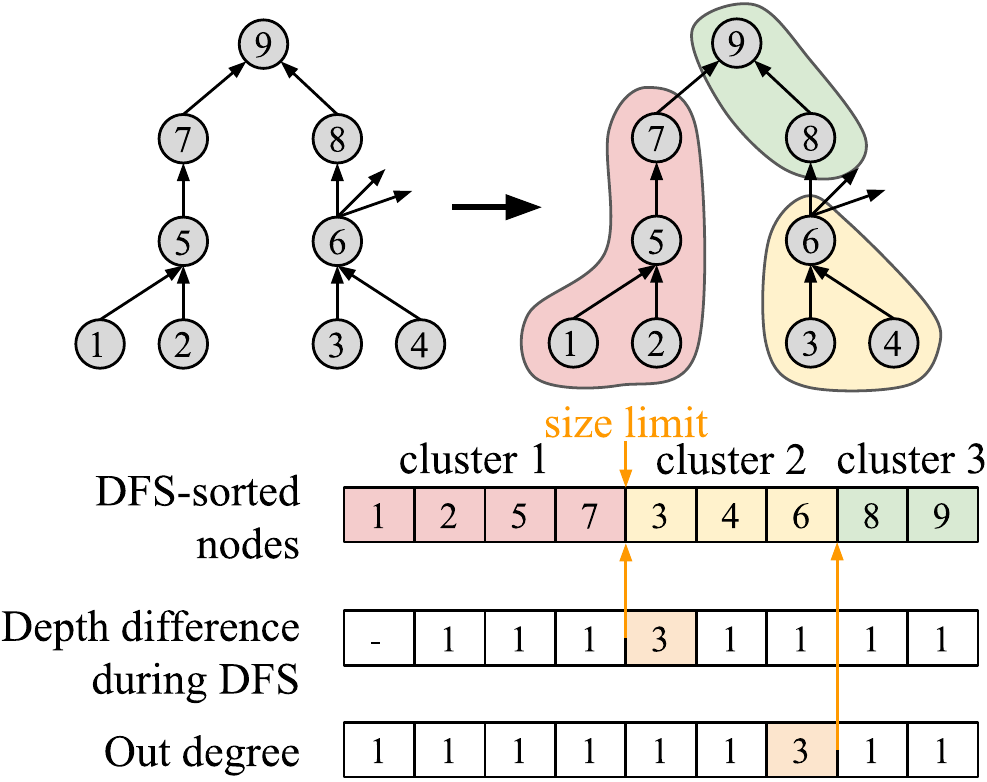}%
\caption{The data structures generated by depth-first traversal for the \textbf{heuristic coarsening step (S3)} \vspace{-5pt}}%
\label{fig:S3}%
\end{figure}

The two-way partitioning model can handle graphs with tens of thousands of nodes and edges in a reasonable time. Hence, the M1 and M2 steps explained earlier are sufficient to generate super layers for such graphs. However, graphs from real applications can have millions of nodes/edges, making the solver runtime prohibitively long. \tool{} uses several techniques to handle the complexity of such large graphs, as shown in the full flow in fig.~\ref{fig:flow_detailed}.
\newline \vspace{-5pt}\\
\textbf{Consider limited layers (S1)}\\
Ideally, all the unmapped nodes of the DAG should be considered for allocation in every iteration of super layer generation. However, during the partitioning for initial \emph{bottom} super layers, it is unlikely that the nodes from the \emph{top} of the graph will be allocated. 
Hence, it is wasteful to consider the entire unmapped graph for the partitioning, and the complexity can be limited by choosing a subset \emph{G}. With this aim, each node in the graph is assigned to a layer number before beginning the generation of super layers, using the "as last as possible" heuristic, such that every node is in one layer below its lowest successor node. \rev{This heuristic only needs a topological sort of graph nodes, which runs in $O(|V| + |E|)$ (i.e., linear in the size of $G$) time}. In every iteration, \tool{} adaptively considers a limited number of layers of the unmapped graph, chosen such that the input graph $G$ for the M1 step has size $\alpha$ (set to 4 in our experiments) times the size of the output super layer in the previous iteration. This automatically chooses an appropriate sized $G$ depending on the parallelism in the DAG. A high $\alpha$ leads to better super layer quality at the expense of partition time, and vice-versa. 
\newline \vspace{-5pt}\\
\textbf{Independent connected components (S2)}\\
During the two-way partitioning in M1 and M2, the current $G$ to partition may not be a fully connected graph, but may have multiple disconnected components. This can simplify the partitioning because disconnected components, by definition, do not have edge crossings. Hence, each component is successively considered as current $G$ and partitioned with the solver independently. Suppose the target for the current recursion is $Y$ threads, then the components are partitioned with a target of $X$ threads, chosen as,
{
\begin{gather}
X= \left\lfloor Y \times \frac{\textrm{size of the current component}}{\textrm{size of all components}}\right\rfloor. \nonumber
\end{gather}
}%
\rev{These independent connected components are discovered using a breadth-first search, running in $O(|V| + |E|)$ (i.e., linear in the size of $G$) time.} 
\newline \vspace{-5pt}\\
\textbf{Heuristic coarsening (S3)}\\
Even with the S1 and S2 techniques, a large graph may have to be two-way partitioned. If the graph is larger than a threshold $\mathit{thresh\_G}$, a list-based heuristic coarsening is used before the partitioning.
The graph nodes are sorted in a list according to the depth-first traversal order as shown in fig.~\ref{fig:S3}, \rev{which runs in $O(|V| + |E|)$ (i.e., linear in the size of $G$) time}. During the traversal, the differences in depth between subsequent nodes are also noted in a list. The third list indicates the out-degree for each node. The node list is then broken into clusters according to the following criteria:
\begin{itemize}
    \item The size of the cluster should be less than a $\mathit{size\_threshold}$. In the example, cluster 1 is stopped at node 7 for the $\mathit{size\_threshold}$ of 4.
    \item The difference in depth among consecutive nodes should be less than a $\mathit{depth\_threshold}$. Eg., the difference in depth for the consecutive nodes 7 and 3 is high, so cluster 1 should be stopped at node 7 according to the depth difference as well. 
    \item Clusters should be stopped at the nodes with out-degree higher than a $\mathit{degree\_threshold}$. Eg., cluster 2 is stopped at node 6 if the $\mathit{degree\_threshold}$ is 2.
\end{itemize}
A coarse graph is constructed with these clusters, which will be used as the current $G$ for two-way partitioning. The clusters are represented as a single node in the coarse graph with a node weight (amount of computation) equal to the sum of all the enclosing node weights. If the resulting graph is very coarse, then it may adversely affect the quality of the subsequent two-way partitioning. As such, the size threshold is chosen according to the current graph size, \rev{resulting in a coarse graph with around 1000 nodes (empirically found to be sufficient for good-quality two-way partitioning)}. The thresholds are as follows,
{
\begin{align*}
& \mathit{size\_threshold} && = \frac{\textrm{size of the current graph to coarsen}}{\textrm{1000}} \\
& \mathit{depth\_threshold} && = log_2(\mathit{size\_threshold}) \\
& \mathit{degree\_threshold} && = 10. 
\end{align*} 
}%
With these three techniques, the \tool{} is able to partition graphs with millions of nodes/edges in seconds. 


\begin{figure*}[!t]
\centering
\includegraphics[trim={0cm 0cm 0cm 0cm} , clip, width=0.97\textwidth]{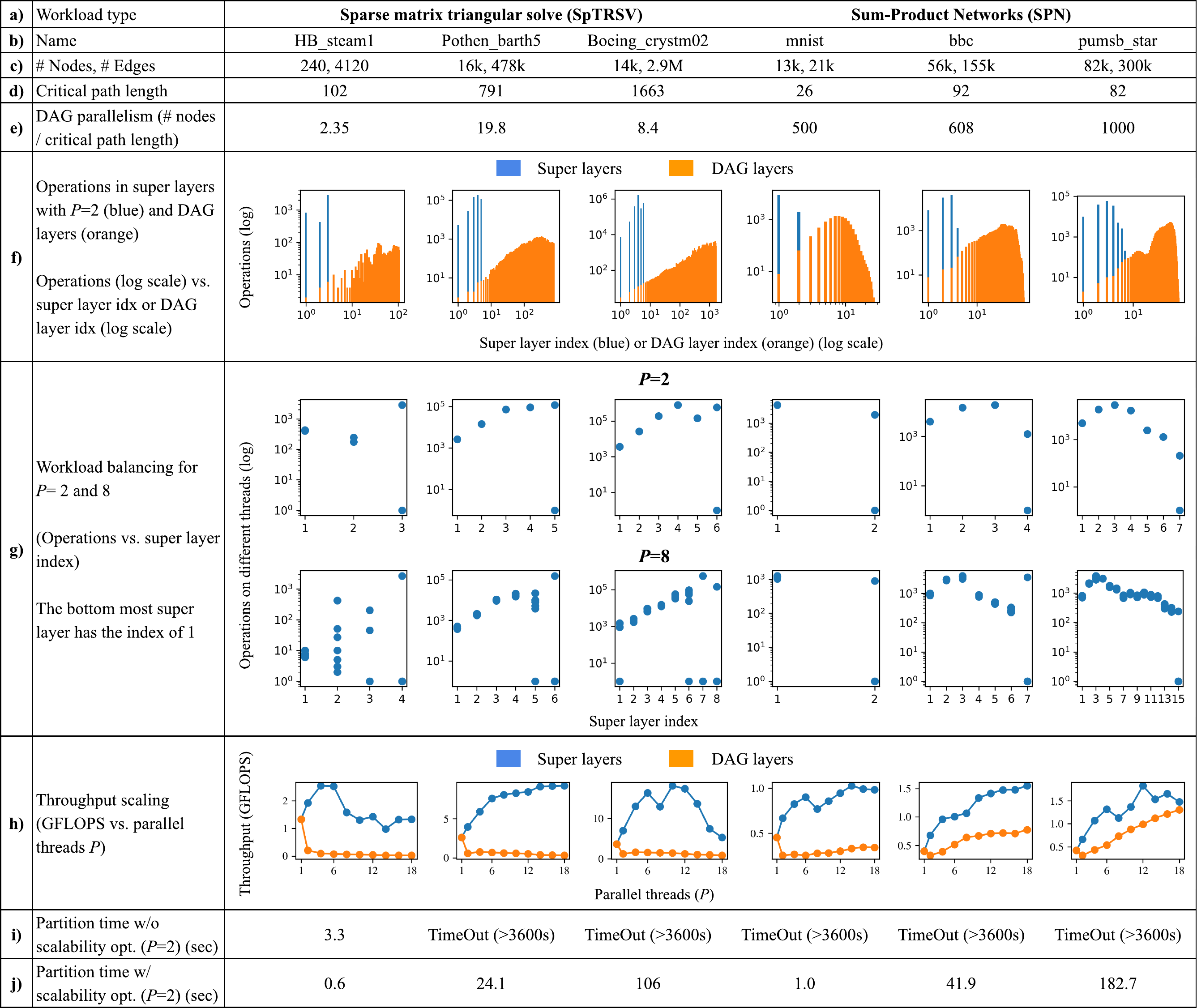}
\caption{\rev{\rtwo{\textbf{Detailed analysis of super layers}. Row~(f) shows that DAGs can be partitioned into a few, large super layers. It also shows the size of the original DAG layers for comparison. Row~(g) shows the workload balancing across threads, in the form of operations in different threads in super layers. Row~(h) shows the throughput scaling with parallel threads, demonstrating the advantage of using super layers versus direct DAG-layer partitioning. Rows~(i) and (j) show the improvement in partitioning time due to scalability techniques S1, S2, and S3.}}}%
\label{fig:results_table}
\end{figure*}
\section{Performance evaluation} \label{sec:experiments}
\subsection{Workloads} \label{sec:workloads}
The multi-threaded CPU performance of \tool{} is evaluated for DAGs from two applications: 1)~Sparse matrix triangular solves of linear algebra, and 2)~Sum-Product Networks of probabilistic machine learning.
\subsubsection{Sparse matrix triangular solves (SpTRSV)} \label{sec:workloads_sptrsv}
Matrix triangular solve is a fundamental operation needed to solve linear system of equations \textbf{L}x=b, where \textbf{L} is a lower (or upper) triangular matrix and the vector x is evaluated for a given vector b \cite{DBLP:journals/actanum/DavisRS16}. This operation has wide-ranging use in engineering simulations, structural analysis, power systems analysis, economics, control, etc. The matrices arising in real-world problems typically turn out to be sparse, i.e. most of the matrix elements are zero. For example, in electric circuit simulations, a matrix can represent the connectivity of a circuit, with rows/columns indicating the circuit nodes and the matrix elements indicating whether two nodes are connected. Since real-world circuit nodes are typically connected to only a few other nodes, such a connectivity matrix will be sparse.
Furthermore, in many applications the sparsity pattern of the matrix remains identical across multiple triangular solves (e.g. the electric circuit structure may remain the same across simulations), providing the opportunity to perform extensive partitioning to improve the runtime performance. 

\rev{The computation of the triangular solve can be represented with a DAG, where each node represents a row, and the node weight is equal to the number of corresponding multiply-accumulate (MAC) operations}. 
If a matrix element \textbf{L}[\textit{i},\textit{j}] is non zero, the DAG contains an edge from node \textit{j} to node \textit{i}, indicating a data dependency. The numbers of DAG nodes and edges are the same as the number of rows and non-zero elements in the matrix, respectively. 

\subsubsection{Sum-Product Networks} \label{sec:workloads_spn}
Sum-Product Networks (SPN, more generally called a probabilistic circuit) \cite{9363463,choi2020probabilistic} is a machine learning workload that can tractably model joint probabilities of random variables. SPNs are often used in combination with neural networks, for various reasoning and perceptual tasks like noise-tolerant image recognition \cite{DBLP:conf/icml/StelznerPK19}, robotic navigation \cite{8967568}, robust human-activity detection \cite{DBLP:conf/nips/OlascoagaMSVB19}, etc. The SPN is a DAG of sum and product operations that can compute complex probabilistic inferences like the marginal and conditional probability of the random variables. The DAG structure is often irregular and remains unchanged during the inference, making it a good candidate for \tool{}.

\rtwo{In addition to these workloads, sparse lower-upper (LU) decomposition and Cholesky decomposition for applications with fixed sparsity structure but changing numerical values (e.g. circuit simulation \mbox{\cite{davis2010algorithm}}) are also good candidates for parallelization with \tool{}.}
\subsection{Experimental setup} \label{sec:experimental_setup}
Experiments are conducted on the matrices from the SuiteSparse matrix collection \cite{DBLP:journals/toms/DavisH11} of real-world applications like power network optimization, structural analysis, computational fluid dynamics, non-linear optimization, economics, robotics, etc. The SPNs from a standard benchmark \cite{DBLP:conf/uai/LiangBB17} are used for evaluation. The super layers generated from the \tool{} are parallelized across multicore CPU threads using OpenMP. The throughput results are averaged over 1000 iterations on 
\rtwo{up to 18 threads on an Intel\textsuperscript{\textregistered} Xeon Gold 6154 CPU with 18 cores, with GCC v4.8.5 compiler, \texttt{-march=native -Ofast} flags, and the thread affinity set as \texttt{KMP\_AFFINITY = granularity = fine,compact,1,0}. 
The experiments with 18 threads are adequate for the target workloads because the workloads have a mean parallelism (quantified as \textit{total nodes in a DAG} / \textit{critical path length}) of 8.6, and 95$\%$ of the DAGs achieve peak throughput with fewer than 18 threads}. 
The caches are \textit{warmed} up by executing the same program before the actual measurement.

\subsection{Analysis of super layers} \label{sec:analysis_of_superlayers}
Several experiments are conducted on workloads of varying sizes to evaluate the different properties of super layers, as summarized in fig. \ref{fig:results_table}.
\subsubsection{How large are the super layers?}%
\tool{} combines nodes from multiple DAG layers to create large super layers. The row~(f) in fig. \ref{fig:results_table} shows the number of multiply-accumulate operations in super layers compared to the original DAG layers. \tool{} manages to compress thousands of DAG layers into tens of super layers, reducing the required synchronization barriers.


\subsubsection{Workload balancing}
Large super layers reduce the number of synchronizations, but good parallel performance demands that the workload is also balanced across threads. The row~(g) in fig.~\ref{fig:results_table} shows the number of operations in each super layer for \Psymbol{} equals to 2 and 8 threads. As seen, \tool{} balances the workload across parallel partitions in a super layer as much as possible. Also note that the granularity of partitions varies across different super layers, depending on the available parallelism in the corresponding DAG region.

\subsubsection{Throughput scaling}
The row~(h) shows the scaling of throughput with parallel threads. Different matrices reach peak throughput with different preferred threads depending on the parallelism in the DAGs. For comparison, the throughput of direct "DAG layer partitioning" is shown in the orange curve. In this partitioning, the nodes of a single DAG layer are executed in parallel across threads, and the threads are synchronized after every DAG layer. The super layers achieve better performance because they need fewer synchronizations. 

\subsubsection{Impact of the scalability techniques}
To see the impact of S1 to S3 scalability techniques on the partitioning time, the partitioning is performed with and without these techniques, with a timeout of 1 hour. Figure~\ref{fig:results_table}(i) and (j)
shows that the tool times-out for the larger graphs without these techniques.
\begin{figure}[!t]
\centering
\includegraphics[trim={0cm 0.57cm 0cm 0cm} , clip, width=\columnwidth]{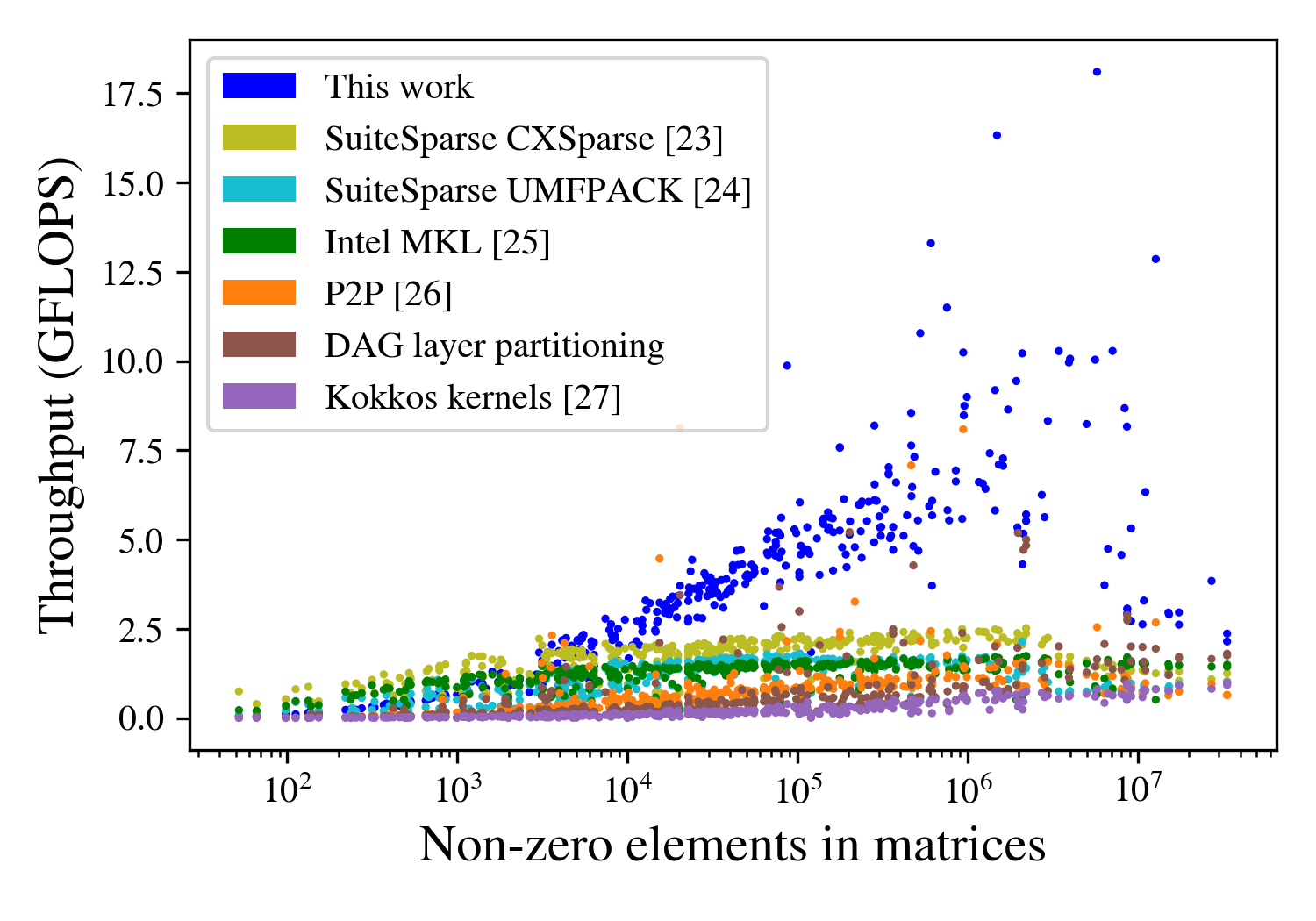}
\caption{\rev{\rtwo{\textbf{Sparse matrix triangular solve performance}. \tool{} achieves a mean speedup of 2.0, 3.3, 3.1, 5.6, 10.8, and 23.7 over SuiteSparse CXSparse, SuiteSparse UMFPACK, Intel MKL, P2P, DAG layer partitioning, and KokkosKernels, respectively.}}}%
\label{fig:scatter}%
\end{figure}%

\subsection{Comparison with state-of-the-art libraries} \label{sec:sota}
\subsubsection{Sparse matrix triangular solves}\label{sec:sota_sptrsv}
As shown in fig. \ref{fig:scatter}, the performance of state-of-the-art libraries is evaluated on the L triangular factor DAG of 370 matrices from the SuiteSparse matrix collection \cite{DBLP:journals/toms/DavisH11}, \rev{with selection criteria that the matrices should be real, square, non-singular, and with $<$50M non-zeroes in the resulting LU factors (to limit the duration of the experiments)}. The experimental details are as follows.
\begin{itemize}[noitemsep, leftmargin=*,topsep=0pt,parsep=0pt,partopsep=0pt]
    \item \textbf{This work}: \tool{} is used to generate super layers with \Psymbol{} set to 2, 4,..., 18 parallel threads. 
    
    \item \rev{\textbf{SuiteSparse}:
Peformance of the CXSparse V3.2.0 \mbox{\cite{davis2006direct}} and the multi-frontal UMFPACK V5.7.9  \mbox{\cite{davis2004algorithm}} algorithms from the widely-used SuiteSparse software collection V5.8.1. The UMFPACK routine is used in Matlab.}

    \item \textbf{Intel MKL}: The \texttt{mkl\_sparse\_s\_trsv()} sparse routine from the Intel Math Kernel Library (MKL) V2021.1 \cite{wang2014intel} is benchmarked for comparison.

\item \textbf{DAG layer partitioning}: This is the same as the orange partitioning in fig. \ref{fig:results_table}, which is a common heuristic approach used in several libraries. 
The nodes of a DAG layer are executed in parallel across threads (which is possible because, by definition, there are no edges within nodes in a layer), and all the threads are synchronized after every layer. The DAG layers are created with the as-late-as-possible (ALAP) heuristic, such that every node is allocated to one layer below its lowest successor node (same as the heuristic used in the S1 step in \S \ref{sec:S1_S2_S3}). 

\item \rev{\textbf{Point-to-Point (P2P)}: Based on \mbox{\cite{park2014sparsifying}} that also uses the DAG layer partitioning, but instead of using global barriers, specialized point-to-point sparsified barriers are used for only the threads involved in dependencies, avoiding unnecessary stalls of the rest of the threads.}

\item \rev{\textbf{KokkosKernels}: Performance of the sparse kernels from the portable Kokkos library V3.4.1 \mbox{\cite{rajamanickam2021kokkos}} is benchmarked.}

\end{itemize}
\noindent \vspace{-8pt}\\
\textbf{Results}: \rev{As shown in \mbox{fig.~\ref{fig:scatter}} the \tool{} outperforms the next best implementation, SuiteSparse CXSparse, on matrices with $>$10k nonzeroes. For smaller matrices, SuiteSparse CXSparse reports the highest performance. Overall, \tool{} achieves mean speedups of 2.0, 3.3, 3.1, 5.6, 10.8, and 23.7 over SuiteSparse CXSparse, Intel MKL, SuiteSparse UMFPACK, P2P, DAG layer partitioning, and KokkosKernels, respectively.} \rtwo{The main contributor for the speedup is fewer synchronization barriers (99$\%$ reduction compared to the DAG layer partitioning).}



\subsubsection{Sum-Product Networks} \label{sec:sota_spn}
\begin{figure}[!t]
\centering
\includegraphics[trim={0cm 0.6cm 0cm 0cm} , clip, width=0.9\columnwidth]{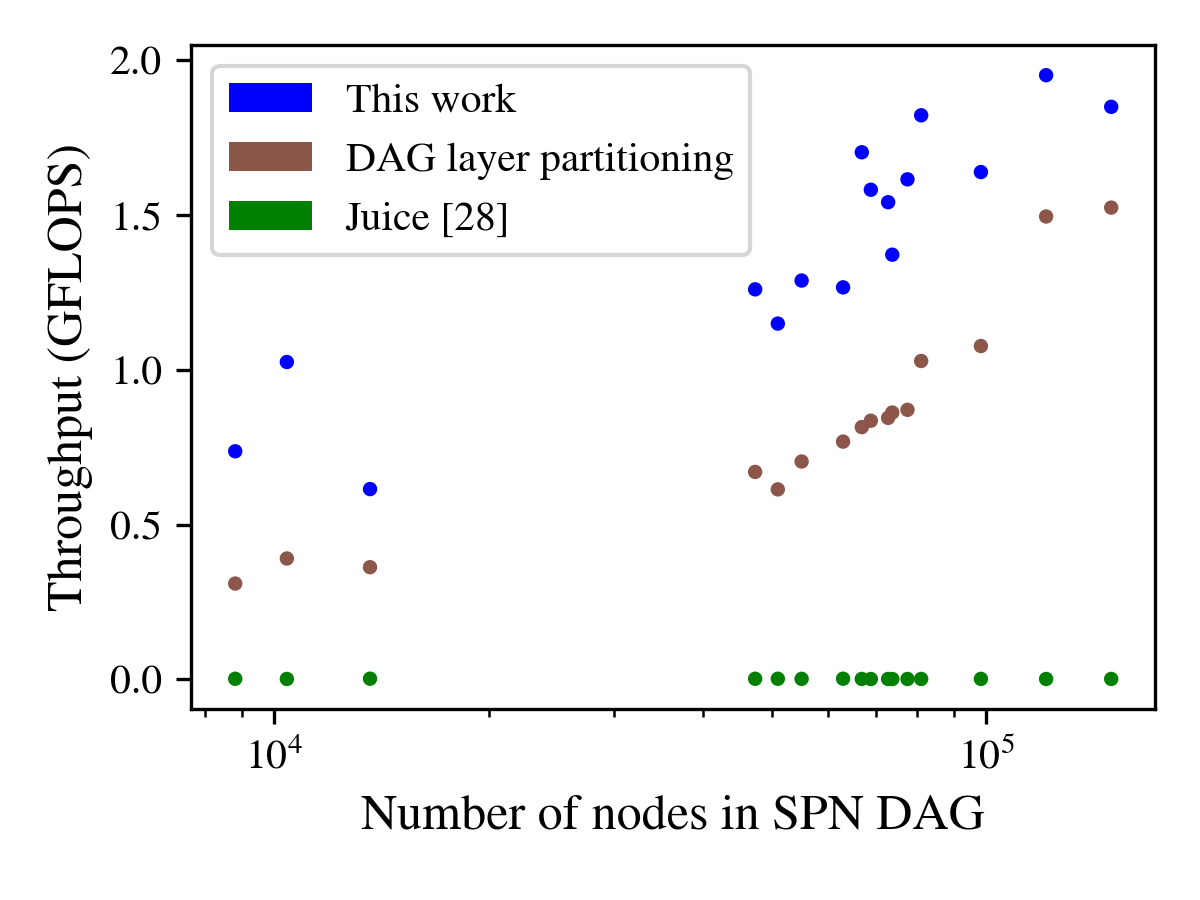}
\caption{\rev{\rtwo{\textbf{Sum-product networks performance}. \tool{} achieves a mean speedup of 1.8$\times$ and 1052$\times$ over DAG layer partitioning and Juice, respectively.}}}%
\label{fig:spn_results}%
\end{figure}

Fig. \ref{fig:spn_results} shows the throughput for 16 SPNs from the standard benchmark \cite{DBLP:conf/uai/LiangBB17} achieved with following implementations. 
\begin{itemize}[noitemsep, leftmargin=*,topsep=0pt,parsep=0pt,partopsep=0pt]

    \item \textbf{This work}: \tool{} is used to generate super layers with \Psymbol{} set to 2, 4,..., 18 parallel threads. 
    
    \item \rev{\textbf{DAG layer partitioning}: Same as the sparse traingular solve experiment in section \S \mbox{\ref{sec:sota_sptrsv}}.}
    
    \item \rev{\textbf{Juice}: Performance of Juice --- a widely-used Julia-based library for SPNs \mbox{\cite{dang2021juice}}.}

\end{itemize}
\noindent \vspace{-8pt}\\
\rev{\textbf{Results}: Juice library, despite of using the layer-based partitioning scheme, achieves significantly lower throughput than the OpenMP-based DAG layer partitioning implementation. \tool{} achieves a mean speed-up of 1.8 and 1052 over DAG layer partitioning and Juice, respectively, \rtwo{due to 88.5$\%$ fewer synchronization barriers.}}

\section{Discussion and related work} \label{sec:related_works}
\subsection{Sparse triangular solves} \label{sec:related_work_sptrsv}
A common approach for parallelizing triangular solves is the DAG layer partitioning method that partitions nodes on each DAG layer directly into \Psymbol{} partitions,
and synchronizes the threads after every layer, first introduced in 
\cite{DBLP:journals/ijhsc/AndersonS89}. This is one of the baselines in the experiments section. This partitioning is quicker than our approach when the sparsity pattern/DAG structure changes for every triangular solve. However, for large matrices, thousands of synchronizing barriers might be needed depending on the critical path length of the DAG, incurring large overhead. As shown in our experiments, 99$\%$ barriers can be avoided by combining multiple DAG layers into super layers, improving the throughput significantly compared to the DAG layer partitioning (fig. \ref{fig:scatter}). \rev{The point-to-point (P2P) barrier approach alleviates some overhead of the global barrier as shown in \mbox{fig. \ref{fig:scatter}}, but still fails to outperform other libraries.}


The works in \cite{DBLP:conf/hipc/PicciauIWKC16} and \cite{DBLP:conf/IEEEpact/HelalACBF19} are the most similar to this paper, but could not be compared due to the unavailability of open-source code. These works have two key conceptual differences with our work:
\begin{enumerate}[leftmargin=*]
    \item Both the papers limits the partition sizes. In \cite{DBLP:conf/hipc/PicciauIWKC16}, the partitions should fit in the local scratchpads of GPU, while in \cite{DBLP:conf/IEEEpact/HelalACBF19}, the size is controlled with a predefined hyperparameter that has to be tuned for every DAG. This hyperparameter remains the same for the entire DAG even when the local parallelism can vary in different parts of the DAG. This is in contrast with our approach of making the partitions as large as possible as long as there are \Psymbol{} parallel partitions. In other words, by explicitly defining the hardware parallelism (via \Psymbol{}), \tool{} automatically adjusts the partition sizes based on the available parallelism in the DAG, reducing the overall number of partitions and global barriers.

    This partition size limit is unnecessary because a global barrier is not needed when partition sizes, e.g., exceed the GPU scratchpad size. Instead, large partitions can be further divided into subpartitions that fit into the scratchpad, and a local barrier (eg. \_\_syncthreads() in CUDA) can be used. 

    \item Both \cite{DBLP:conf/hipc/PicciauIWKC16} and \cite{DBLP:conf/IEEEpact/HelalACBF19} develop a partitioning heuristic, while \tool{} uses constrained optimization and leverages a state-of-the-art solver, which often achieve better solutions than custom heuristics.
\end{enumerate}%
\subsection{Sum-Product Networks} \label{sec:related_work_spn}
The DAG layer partitioning heuristic is used for SPN for CPU, GPU and a custom ASIC parallelization \cite{dang2021juice, DBLP:conf/date/ShahOMV20, dadu2019towards}. \rev{Our experiments show a speedup of 1052$\times$ over Juice \mbox{\cite{dang2021juice}}}. Libraries based on Tensorflow have also been developed for coarse-grained SPNs with explicit regular structures \cite{DBLP:conf/iros/PronobisR17, Molina2019SPFlow}. However, such a Tensorflow-based approach is not useful for fine-grained, irregular SPNs due to the overheads of kernel launch etc. \tool{} does not assume any regularity in SPN structure.
\subsection{Graph partitioning} \label{sec:related_work_graph_part}
As explained in section \S \ref{sec:graph_opt}, \tool{} essentially needs to solve \Psymbol{}-way independent partitioning of a directed graph while ensuring that resulting partitions are acyclic and balanced. In general, graph partitioning is known to be NP-complete \cite{DBLP:journals/tcs/Feldmann13, DBLP:journals/siamsc/KarypisK98} and is a widely studied problem. Several partitioning algorithms have been developed \cite{9115834,  DBLP:series/lncs/BulucMSS016, 
bichot2013graph
}, but they only focus on undirected graphs intending to reduce the edge crossings within balanced partitions, while ignoring the edge direction. As a result, the popular undirected partitioning software like JOSTLE \cite{walshaw2007jostle} and METIS \cite{DBLP:journals/siamsc/KarypisK98} cannot be used for acyclic partitioning.

The acyclic partitioning of DAGs is also shown to be NP-complete like the undirected version of the problem \cite{DBLP:journals/heuristics/MoreiraPS20}. In recent years, several works have been proposed to tackle the problem \cite{
DBLP:journals/siamsc/HerrmannOUKC19, DBLP:journals/heuristics/MoreiraPS20, DBLP:conf/wea/MoreiraPS17, DBLP:conf/dac/CongLB94}. However, these works do not focus on parallelism. The resulting partitions would be acyclic, well-balanced, with minimal edge crossings, but can end up being completely sequential, i.e. only one partition can be executed at a time. This stems from the fact that the usual objective of minimizing edge crossings does not guarantee parallelism. Hence, these methods are not suitable for parallelizing a DAG execution over multiple threads.

Several algorithms have been proposed for scheduling DAGs \cite{DBLP:journals/peerj-cs/BramasK20, 
DBLP:conf/ipps/OzkayaBUHC19, 8301529 
}, which use either list-based or clustering-based scheduling heuristics, while \tool{} takes a different approach of modeling the core routine of the tool as a constrained-optimization problem, allowing the use of open-source solvers.
The constrained optimization-based approach is explored in \cite{valouxis2013dag, 6889114}. \rev{However, both these works use several simplifying assumptions which do not hold for a CPU multithreaded execution, for example, assumptions like (1) all the threads are synchronous, (2) execution and communication latencies are fixed and predefined, and (3) a node execution can be launched precisely in a given cycle. As such, these models cannot generate valid schedules for asynchronous multithreaded CPU execution. Hence, to the best of our knowledge, ours is the first work employing constrained optimization for multithreaded CPU scheduling of irregular DAGs.}
\subsection{Suggestions for future work} \label{sec:future_work}
The two-way partitioning model in section \S \ref{sec:M1} can be extended to \Psymbol{}-way partitions by simply allowing the \emph{PART} variable to take values in integers [0,P] instead of [0,2]. This would find \Psymbol{} partitions of a super layer directly, avoiding recursive two-way partitioning. However, allowing values in [0,P] adds symmetries in the search space since two partitions can be switched without breaking any constraint. These symmetries prohibitively increase the run time of the solver since it keeps finding (or refuting) candidates that are just permutations of an earlier solution (or non-solution). The usual technique of symmetry breaking using lexicographical ordering constraint on the \emph{PART} variable is not sufficient because it does not break all the symmetries. A recent paper \cite{DBLP:journals/constraints/CodishMPS19} proposes stricter symmetry breaking constraints for several graph optimization problems, which we believe can also be applied to this problem. We leave this as a future improvement to this work. 








\section{Conclusion} \label{sec:conclusion}
This paper describes \tool{}, a tool developed to efficiently parallelize sparse, irregular graph workloads on parallel compute threads. Graphs are decomposed into super layers with 
\Psymbol{} parallel partitions, using multiple recursions of two-way partitioning of subgraphs. The two-way partitioning problem is modeled as an optimization problem with the Minizinc constraint-modeling language and solved with the open Google OR-Tools solver. The full flow of \tool{} also contains steps for workload balancing and scalability techniques to handle large graphs. The resulting performance of this super layer-based partitioning is benchmarked for sparse matrix triangular solves and sum-product networks, respectively achieving a speed-up of 2.0$\times$ and 1.8$\times$ over the best existing libraries. Thus, \tool{} demonstrates that constrained optimization is effective for the parallelization of large graph workloads.
\ifCLASSOPTIONcompsoc
  \section*{Acknowledgments}
\else
 \section*{Acknowledgment}
\fi
\rtwo{This work has been supported by the European Union's European Research Council (ERC) project Re-SENSE under grant agreement ERC-2016-STG-715037, and we acknowledge support from Intel.}


\ifCLASSOPTIONcaptionsoff
  \newpage
\fi



\bibliographystyle{IEEEtran}
\bibliography{IEEEabrv,ref_wo_url.bib}

\begin{IEEEbiography}[{\includegraphics[width=1in,height=1.25in,clip,trim={4cm 17cm 4cm 0cm},keepaspectratio]{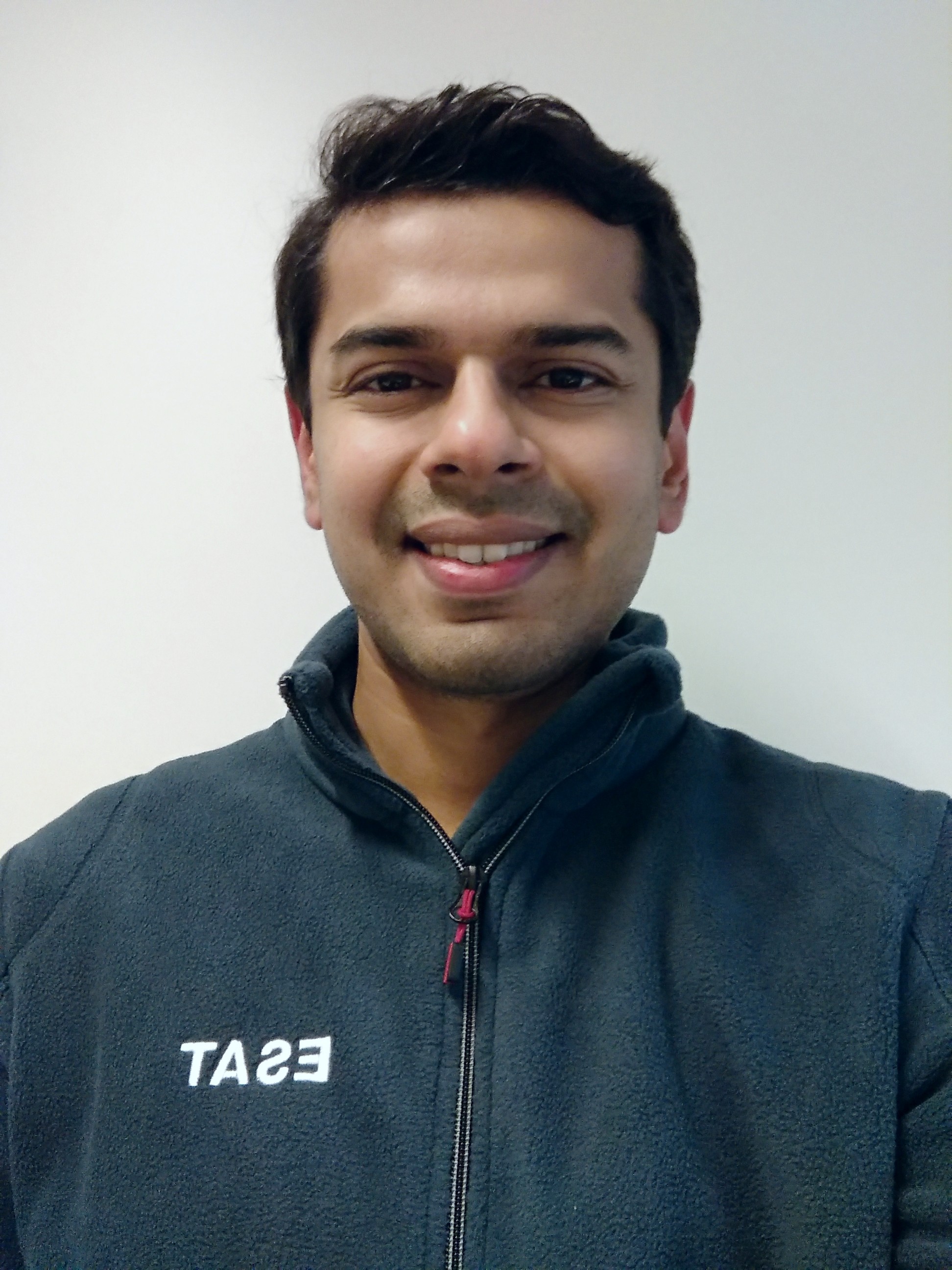}}]{Nimish Shah}
is pursuing his Ph.D. degree at the MICAS laboratories of the EE department of KU Leuven, Belgium. His research focuses on hardware-software co-design, embedded machine learning, irregular graph processing, approximate computing, and low-power digital VLSI. He received an M.Tech. degree in Electronic Systems Engineering from the Indian Institute of Science, Bangalore, in 2016. In 2016-17, he worked with Nvidia, Bangalore, where he was involved in energy-efficient memory compression hardware for GPU. Nimish is a recipient of the departmental Gold Medal for excellence in master’s studies at IISc.
\end{IEEEbiography}
\vspace{-0.5cm}

\begin{IEEEbiography}[{\includegraphics[width=1in,height=1.25in,clip, keepaspectratio]{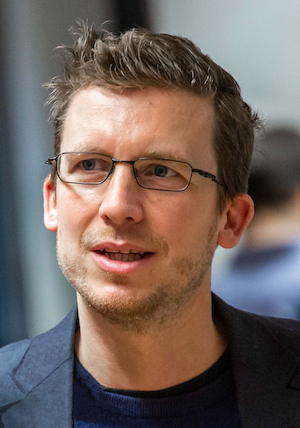}}]{Wannes Meert}
received his degrees of Master of Electrotechnical Engineering, Micro-electronics (2005), Master of Artificial Intelligence (2006) and Ph.D. in Computer Science (2011) from KU Leuven. He is an IOF research manager in the DTAI section at KU Leuven. His work is focused on applying machine learning, artificial intelligence and anomaly detection technology to industrial application domains with various industrial and academic partners.
\end{IEEEbiography}
\vspace{-0.5cm}
\begin{IEEEbiography}[{\includegraphics[width=1in,height=1.25in,clip,keepaspectratio]{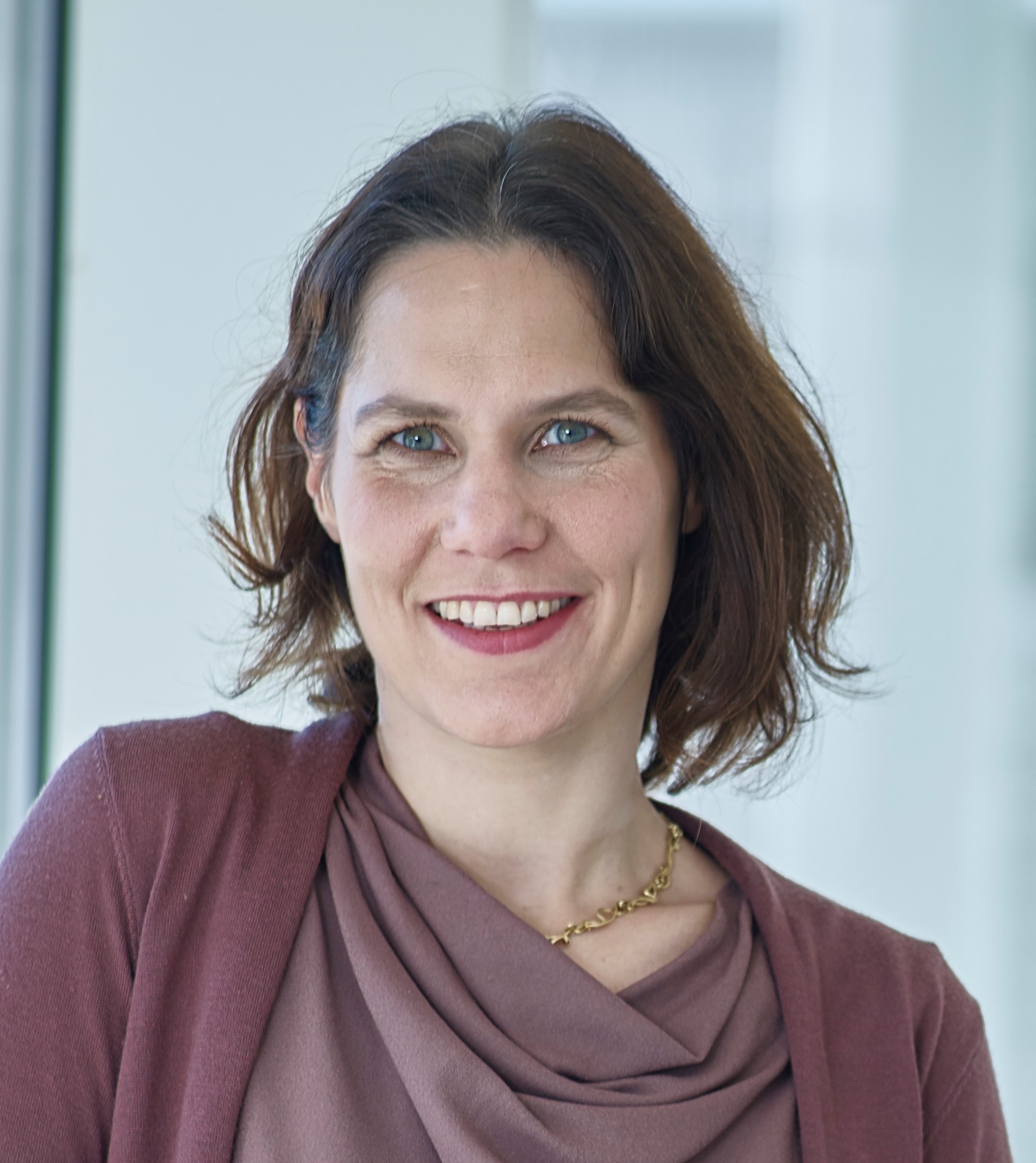}}]{Marian Verhelst}
is a full professor at the MICAS laboratories of the EE Department of KU Leuven. Her research focuses on embedded machine learning, hardware accelerators, HW-algorithm co-design and low-power edge processing. Before that, she received a PhD from KU Leuven in 2008, was a visiting scholar at the BWRC of UC Berkeley in the summer of 2005, and worked as a research scientist at Intel Labs, Hillsboro OR from 2008 till 2011. Marian is a topic chair of the DATE and ISSCC executive committees, TPC member of VLSI and ESSCIRC  and was the chair of tinyML2021 and TPC co-chair of AICAS2020. Marian is an IEEE SSCS Distinguished Lecturer, was a member of the Young Academy of Belgium, an associate editor for TVLSI, TCAS-II and JSSC and a member of the STEM advisory committee to the Flemish Government. Marian currently holds a prestigious ERC Starting Grant from the European Union, was the laureate of the Royal Academy of Belgium in 2016, and received the André Mischke YAE Prize for Science and Policy in 2021.
\end{IEEEbiography}

\clearpage

\newpage
\thispagestyle{empty} 
\onecolumn
\appendices
\section{\rev{ \tool{} algorithm}} \label{sec:appendix_a}
This appendix describes the different algorithms used in \tool{}.
\begin{itemize}
    \item \textbf{Algorithm \ref{alg:full}}: The top-level algorithm of \tool{}.
    \item \textbf{Algorithm \ref{alg:GenDAGLayers}}: Maps nodes to DAG layers according to the "As last as possible" (ALAP) heuristic. These DAG layers are required for the S1 scalability technique. 
    \item \textbf{Algorithm \ref{alg:S1}}: The S1 scalability algorithm of limiting layers as described in section \S \ref{sec:S1_S2_S3}.
    \item \textbf{Algorithm \ref{alg:M1_S2_S3}}: The recursive two-way partitioning step M1 (section \S \ref{sec:M1}) which in turn uses the scalability techniques S2 and S3. The S2 step (section \S \ref{sec:S1_S2_S3}) of generating DAG components is readily available in Python's NetworkX library.
    \item \textbf{Algorithm \ref{alg:S3}}: Describes the S3 step of heuristic coarsening, as explained in section \S \ref{sec:S1_S2_S3}
    \item \textbf{Algorithm \ref{alg:M2}}: The workload balancing step explained in section \S \ref{sec:M2}
\end{itemize}
The two-way partitioning model used in algo. \ref{alg:M1_S2_S3} and \ref{alg:M2} is described in appendix \textbf{\ref{sec:minizinc_code}}.

\newcommand\mi[1]{\mathit{#1}}

\newcommand\mycommfont[1]{\footnotesize\ttfamily\textcolor{blue}{#1}}
\SetCommentSty{mycommfont}

\SetKwComment{tcp}{ \quad $\triangleright$ }{}%
\SetKwComment{tcc}{$\triangleright$ }{}%

\noindent\begin{minipage}[t]{0.99\textwidth}

\begin{algorithm}[H]
\DontPrintSemicolon 
\KwIn{$V_{\mathit{full}}$, $E_{\mathit{full}}$: set of nodes and edges of target DAG,\newline
$\mathit{node\_w}$: node weights, \newline
$\mathit{nThreads}$: number of threads \newline
}
\KwOut{$\mathcal{D}_{\mathit{th}}$: Dictionary mapping every node to a thread,\newline
$\mathcal{D}_{\mathit{sl}}$: Dictionary mapping every node to a superlayer\newline}

$G_{\mathit{full}} \gets$ networkx.DiGraph$\langle V_{\mathit{full}}, E_{\mathit{full}} \rangle$ \tcp{Graph datastructure from the NetworkX Python library }

$\mathit{DAG\_layers} \gets$ GenDAGLayers$\langle G_{\mathit{full}}\rangle$  \tcp{Algo \ref{alg:GenDAGLayers}}

$V_{\mathit{unmapped}}$ $\gets$ $V_{\mathit{full}}$ \;

$\mathcal{D}_{\mathit{th}} \gets \{\}$ \tcp{Empty dictionary}

$\mathcal{D}_{\mathit{sl}} \gets \{\}$ \tcp{Empty dictionary}

$\mathit{last\_mapped\_count} \gets 0$ \tcp{Number of nodes mapped in the previous iteration}

$\mathit{sl} \gets 0$ \tcp{Current superlayer index}
\;
\While (
) {$V_{\mathit{unmapped}} \neq \varnothing$} 
{
    $V_{a}$ $\gets$ S1$\langle  \mathit{last\_mapped\_count}, \mathit{DAG\_layers}\rangle$ \tcp{Algo \ref{alg:S1}}
    
    \;
    $\mathit{mapped\_}\mathcal{D}_{\mathit{th}}
    \gets$ M1\_S2\_S3$\langle G_{\mathit{full}}, V_a, 
    \mathcal{D}_{\mathit{th}}, \mathit{nThreads} \rangle$ \tcp{Algo \ref{alg:M1_S2_S3}}
    
    
    $\mathit{mapped\_}\mathcal{D}_{\mathit{th}}
    \gets$ M2$\langle
    G_{\mathit{full}},
    \mathit{mapped\_}\mathcal{D}_{\mathit{th}}, \mathcal{D}_{\mathit{th}}, \mathit{nThreads}
    \rangle$ \tcp{Algo \ref{alg:M2}}
    
    \;
    $\mathit{curr\_}V_{\mathit{done}} \gets
    \mathit{mapped\_}\mathcal{D}_{\mathit{th}}$.keys() \tcp{Dictionary keys() are the mapped nodes}
    
    
    $V_{\mathit{unmapped}} \gets V_{\mathit{unmapped}} \backslash \mathit{curr\_}V_{\mathit{done}}$
    
    \;
    $\mathcal{D}_{\mathit{th}}$.update$\langle \mathit{mapped\_}\mathcal{D}_{\mathit{th}} \rangle$
    
    $\mathit{mapped\_}\mathcal{D}_{\mathit{sl}} \gets$ \{ $v$ : $sl$ for v in $\mathit{curr\_}V_{\mathit{done}}$ \}
    
    $\mathcal{D}_{\mathit{sl}}$.update$\langle \mathit{mapped\_}\mathcal{D}_{\mathit{sl}} \rangle$\;
    \;
    
    $\mi{DAG\_layers}$.remove($curr\_V_{done}$)
    
    $\mathit{last\_mapped\_count} \gets $ len$(\mathit{curr\_}V_{\mathit{done}})$
    
    $sl$++
}
\Return{$\mathcal{D}_{\mathit{th}}$, $\mathcal{D}_{\mathit{sl}}$}
\caption{\tool{} $\langle$ $V_{\mathit{full}}$, $E_{\mathit{full}}$,
$\mathit{node\_w}$,
$\mathit{nThreads}$ $\rangle$}
\label{alg:full}

\end{algorithm}  


  \end{minipage} 

\clearpage
\thispagestyle{empty} 
\noindent \begin{minipage}[t]{0.99\textwidth}

\begin{algorithm}[H]
\DontPrintSemicolon 
\tcc{Generate DAG layers with "As late as possible" (ALAP) heuristic}
\KwIn{$G_{\mathit{full}}$: Full graph (in Python's NetworkX datastructure)\newline
}
\KwOut{$\mi{DAG\_layers}$: Layers of DAGs in the form of list of sets of nodes \newline
}

$\mi{topo\_order} \gets$ NetworkX.topological\_sort($G_{full}$)

$\mi{DAG\_layers} \gets [set() \; for \; l \; in$ range( NetworkX.dag\_longest\_path($G_\mi{full}$) ) $]$

\For{$v \mathrm{\; in \;} \mi{topo\_order}$}
{
    $\mi{layer\_id} \gets 0$
    
    \If{$G_\mi{full}\mathrm{.successors(v)}$}
    {
        $\mi{layer\_id} \gets$ max(layer\_id of $G_\mi{full}$.successors()) + 1
    }
    
    $\mi{DAG\_layers}[\mi{layer\_id}]$.add($v$)
}
$\mi{DAG\_layers}$.reverse() \tcp{bottom layer at index 0}

\Return{$\mi{DAG\_layers}$}

\caption{GenDAGLayers $\langle G_\mi{full} \rangle$}

\label{alg:GenDAGLayers}

\end{algorithm}  


  \end{minipage} 

\noindent \begin{minipage}[t]{0.99\textwidth}

\begin{algorithm}[H]
\DontPrintSemicolon 
\tcc{Limit layers (section \S \ref{sec:S1_S2_S3})}
\KwIn{$\mi{last\_mapped\_count}$: Number of nodes mapped in the last superlayer,\newline
    $\mi{DAG\_layers}$: Remaining unmapped layers \newline
}
\KwOut{$V_a$: Nodes to consider for current superlayer\newline
    }

$V_a \gets $ set()

\For{$\mi{layer} \mathrm{\; in \;} \mi{DAG\_layers}$}
{
    $V_a \gets V_a \cup \mi{layer}$
    
    \If(\tcp*[h]{Limit layers depending on the size of previous super layer. $\alpha$ is set to 4 in our experiments}){$len(V_a) > \alpha \times \mi{last\_mapped\_count}$}
    {   
        break
    }
}
\Return{$V_a$}

\caption{S1  $\langle G, \mi{node\_w} \rangle$}
\label{alg:S1}

\end{algorithm}  


  \end{minipage} 
\newline
\clearpage
\thispagestyle{empty} 
\noindent\begin{minipage}[t]{0.99\textwidth}

\begin{algorithm}[H]
\DontPrintSemicolon 
\tcc{Recursive two-way partitioning (section \S \ref{sec:M1}) with scalability techniques}
\KwIn{
    $V_a$: Target nodes, \newline
    $G_{\mathit{full}}$: Full graph (in Python's NetworkX datastructure),\newline
    $\mathit{node\_w}$: node weights, \newline
    $\mathcal{D}_{\mathit{th}}$: Dictionary containing mapping of previously generated superlayers,\newline
    $Y$: set of target threads \newline
}
\KwOut{$\mathit{curr\_}\mathcal{D}_{\mathit{th}}$: Partitions in the form of a dictionary mapping nodes to threads\newline
    }

$\mathit{curr\_}\mathcal{D}_{\mathit{th}} \gets$ \{\}

\;
\If(\tcp*[h]{base case of recursion}){len($Y$) = 1 }
{
    $th \gets Y$.pop()
    
    \For{$v \mathrm{\; in \;} V_a$}
    {   
        $\mathit{curr\_}\mathcal{D}_{\mathit{th}}[v] \gets th$
    }
    \Return{$\mathit{curr\_}\mathcal{D}_{\mathit{th}}$}
}
\;
$\mi{subG} \gets G_{\mi{full}}$.subgraph($V_a$)

$\mi{components} \gets$ NetworkX.weekly\_connected\_components($subG$) \tcp{S2 step}

\For{$V \mathrm{\; in \;} \mi{components}$}
{
    \If (\tcp*[h]{Perform step S3}) {$len(V) > \mi{thresh\_G}$}
    {
        $\mi{currG} \gets G_{\mi{full}}$.subgraph($V$)

        $\mi{coarse\_G, coarse\_node\_w} \gets $ S3$\langle \mi{currG} ,  \mathit{node\_w} \rangle$ \tcp{Algo \ref{alg:S3}}
    
        $V \gets \mi{coarse\_G}$.nodes()
    
        $\mathit{curr\_node\_w} \gets \mathit{coarse\_node\_w}$
    }
    \Else
    {
        $\mathit{curr\_node\_w} \gets \mathit{node\_w}$
    }
    
    \;
    
    $nX \gets len(Y) \times len(V) \; / \; len(V_a)$
    
    $X1 \gets nX/2$ threads from $Y$ \tcp{Target threads for $\mi{PART\_1}$}
    
    $Y$.remove($X1$)

    $X2 \gets nX/2$ threads from $Y$ \tcp{Target threads for $\mi{PART\_2}$}

    $Y$.remove($X2$)
    
    \;
    
    $\mi{PART\_1}, \mi{PART\_2} \gets$ TwoWayPartition $\langle $V$, \mathit{curr\_node\_w}, X1, X2, G_{\mathit{full}}, \mathcal{D}_{\mathit{th}} \rangle$ \tcp{Appendix \ref{sec:minizinc_code}}
    \;
    
    \tcc{If S3 (line 14) used, transform $\mi{PART\_1}$ and $\mi{PART\_2}$ from coarse to original nodes}
    \;
    
    $\mathit{curr\_}\mathcal{D}_{\mathit{th}}\_1 \gets$ M1\_S2\_S3  $\langle \mi{PART\_1}, G_{\mathit{full}}, \mi{node\_w},
    \mathcal{D}_{\mathit{th}}, \mathit{X1} \rangle$ \tcp{Recurse on each partition}
    
    $\mathit{curr\_}\mathcal{D}_{\mathit{th}}\_2 \gets$ M1\_S2\_S3  $\langle \mi{PART\_2}, G_{\mathit{full}}, \mi{node\_w},
    \mathcal{D}_{\mathit{th}}, \mathit{X2} \rangle$ 

    $\mathit{curr\_}\mathcal{D}_{\mathit{th}}$.update($\mathit{curr\_}\mathcal{D}_{\mathit{th}}\_1$)
    
    $\mathit{curr\_}\mathcal{D}_{\mathit{th}}$.update($\mathit{curr\_}\mathcal{D}_{\mathit{th}}\_2$)

}
\Return{$\mathit{curr\_}\mathcal{D}_{\mathit{th}}$}

\caption{M1\_S2\_S3  $\langle V_a,      G_{\mathit{full}},
    \mi{node\_w},
    \mathcal{D}_{\mathit{th}}, \mathit{Y} \rangle$
    }
\label{alg:M1_S2_S3}

\end{algorithm}  


  \end{minipage} 

\clearpage
\thispagestyle{empty} 
\noindent\begin{minipage}[t]{0.99\textwidth}

\begin{algorithm}[H]
\DontPrintSemicolon 
\tcc{Heuristic coarsening (section \ref{sec:S1_S2_S3})}
\KwIn{$G$: Target graph (in Python's NetworkX datastructure),\newline
    $\mathit{node\_w}$: node weights, \newline
    }
\KwOut{$\mi{coarse\_}G$: Coarse graph,\newline
    $\mathit{coarse\_node\_w}$: coarse node weights, \newline
    }

\tcc{Generate $\mi{dfs\_node\_ls}$ and $\mi{depth\_diff\_ls}$ datastructures}

$\mi{root\_nodes} \gets $ nodes in $G$ with no successors

$\mi{done\_nodes} \gets \varnothing$

$\mi{depth\_diff} \gets 0$

$\mi{depth\_diff\_ls} \gets []$

$\mi{dfs\_node\_ls} \gets []$

\While{$\mi{root\_nodes}$}
{
    $\mi{stack} \gets [\mi{root\_nodes}$.pop()$]$
    
    \While{$\mi{stack}$}
    {
        $\mi{curr\_v} \gets stack[-1]$ \tcp{Last element from the stack}
        
        $\mi{depth\_diff}$++
        
        $\mi{next\_ls} \gets [v |v \in G$.predecessors($\mi{curr\_v}$) and $v \notin \mi{done\_nodes}]$
        \;
        
        \If{$\mi{next\_ls}$}
        {   
            $\mi{stack}$.extend($next\_ls$)
        }
        \Else{
            $\mi{done\_nodes}$.add($\mi{curr\_v}$)
            
            $\mi{node\_ls}$.append($\mi{curr\_v}$)
            
            $\mi{depth\_diff\_ls}$.append($\mi{depth\_diff}$)
            
            $\mi{depth\_diff} \gets 0$
            
            $\mi{stack}$.pop() \tcp{Pop $\mi{curr\_v}$ from the stack}
        }
    }
}
\;
\tcc{Generate coarse nodes}

$\mi{map\_coarse\_node\_to\_node\_set} \gets $ \{\}

$\mi{coarse\_node\_w} \gets $ \{\}

$\mi{coarse\_nodes\_id} \gets 0$

$\mi{curr\_node\_set} \gets $ set()

$\mi{curr\_node\_w} \gets 0$

$\mi{size\_threshold} \gets \frac{G.\mathrm{size()}}{1000}$ \tcp{$\mi{coarse\_G}$ will have around 1000 coarse nodes}

$\mi{depth\_threshold} \gets  log(\mi{size\_threshold})$

$\mi{degree\_threshold} \gets 10$

\For{$(i, v) \mathrm{\; in \; enumerate}(\mi{dfs\_node\_ls}$)}
{   
    
    \If(\tcp*[h]{Create a coarse node}){(len($\mi{curr\_node\_set}$) $> \mi{size\_threshold}$) \textbf{or} \newline ($\mi{depth\_diff\_ls}[i] > \mi{depth\_threshold}$) \textbf{or} \newline ($G.out\_degree(v) > \mi{degree\_threshold}$)}
    {
        $\mi{map\_coarse\_node\_to\_node\_set}[\mi{coarse\_node\_id}] \gets \mi{curr\_node\_set}$

        $\mi{curr\_node\_set} \gets set()$
        
        \;
        $\mi{coarse\_node\_w}[\mi{coarse\_node\_id}] \gets \mi{curr\_node\_w}$
        
        $\mi{curr\_node\_w} \gets 0$
        
        $\mi{coarse\_node\_id}$++

    }
    $\mi{curr\_node\_set}$.add($v$)
    
    $\mi{curr\_node\_w}$ += $\mi{node\_w}[v]$    
}
$\mi{coarse\_G} \gets$ Generate python NetworkX graph from $\mi{map\_coarse\_node\_to\_node\_set}$ and $G$

\Return{$\mi{coarse\_G}, \mi{coarse\_node\_w}$}

\caption{S3  $\langle G, \mi{node\_w} \rangle$}
\label{alg:S3}

\end{algorithm}  


  \end{minipage} 

\clearpage
\thispagestyle{empty} 
\noindent\begin{minipage}[t]{0.99\textwidth}

\begin{algorithm}[H]
\DontPrintSemicolon 
\tcc{Balance partitions (section \S \ref{sec:M2})}
\KwIn{$G_{\mathit{full}}$: Full graph (in Python's NetworkX datastructure),\newline
    $\mathit{node\_w}$: node weights, \newline
    $\mathit{imbalanced\_}\mathcal{D}_{\mathit{th}}$: Imbalanced partitions, represented as a dictionary mapping nodes to threads,\newline
    $\mathcal{D}_{\mathit{th}}$: Dictionary containing mapping of previously generated superlayers,\newline
    $\mathit{nThreads}$: number of threads \newline
}
\KwOut{$\mathit{balanced\_}\mathcal{D}_{\mathit{th}}$: Balanced partitions, represented as a dictionary mapping nodes to threads\newline
}
$\mi{partition\_ls} \gets$ [set\ ([ $v$ $|$   $\mi{imbalanced\_}\mathcal{D}_{\mi{th}}[v] \; = \; th$ ]) $|$ $\forall th \; \in \;$[$0, \mi{nThreads})$]\tcp{list of partitions (which are represented as sets)}

$\mi{thread\_pool} \gets \mi{nThreads}$

\While{$\mathrm{len}(\mi{thread\_pool}$) $> \; 1$}
{
    $th\_L \gets$ Thread with largest partition among $\mi{thread\_pool}$

    $th\_S \gets$ Thread with smallest partition among $\mi{thread\_pool}$
    
    $L \gets \mi{partition\_ls}[th\_L]$ \tcp{currently largest partition}

    $S \gets \mi{partition\_ls}[th\_S]$ \tcp{currently smallest partition}
    
    \;
    $\mi{Combined} \gets S \cup L$
    
    $L', S' \gets $ TwoWayPartition $\langle \mi{Combined}, \mathit{node\_w}, \{th\_L\}, \{th\_S\}, G_{\mathit{full}}, \mathcal{D}_{\mathit{th}} \rangle$ \tcp{Appendix \ref{sec:minizinc_code}}
    \;
    \If(\tcp*[h]{New partitions are more balanced}){$\mathrm{min}(\mathrm{len}(L',S') > \mathrm{len}(S)$}  
    {
        $\mi{partition\_ls}[th\_L] \gets L'$
        
        $\mi{partition\_ls}[th\_S] \gets S'$
    }
    \Else(\tcp*[h]{L is not divisible further due to lack of parallelism. Do not try again.})
    {
        $\mi{thread\_pool}$.remove($L$)
    }
    
}
\For{$(th, \mi{partition}) \mathrm{\; in \; enumerate}(\mi{partition\_ls})$}
{
    \For{$v \mathrm{\; in \;} \mi{partition}$}
    {   
        $\mathit{balanced\_}\mathcal{D}_{\mathit{th}}[v] \gets th$
    }
}

\Return{$\mathit{balanced\_}\mathcal{D}_{\mathit{th}}$}

\caption{M2  $\langle G_{\mathit{full}},       \mi{node\_w},
    \mathit{imbalanced\_}\mathcal{D}_{\mathit{th}}, \mathcal{D}_{\mathit{th}}, \mathit{nThreads} \rangle$}
\label{alg:M2}

\end{algorithm}  


  \end{minipage} 

\clearpage
\thispagestyle{empty} 
\section{\rev{Two-way partitioning model}} \label{sec:minizinc_code}

The code listing \ref{lst:minizinc_code} is the minizinc-based optimization model for two-way partitioning of graphs, as described in section \S \ref{sec:M1}. A python wrapper to this minizinc model generates the input parameters like $V_{in},E_{in},\mi{PART\_{in}}$ from the graph structure and mapping of previous superlayers. The code listing \ref{lst:minizinc_dzn_example} describes the input parameters to the model for the example shown in fig. \ref{fig:example}.\newline

\noindent\hspace{0.05\textwidth}\begin{minipage}{.9\textwidth}

\begin{lstlisting} [
language=minizinc,
caption= {Minizinc code for the two-way partitioning optimization model described in section \S \ref{sec:M1}},
label= {lst:minizinc_code}
]
%%%%% input parameters %%%%
int: n_V; % number of nodes in graph G
set of int: V = 1..n_V;

int: max_node_w; % maximum node weight in current G
array[V] of 1..max_node_w: node_w;

int: n_E;
array[1..n_E, 1..2] of int: E; % edges (an array (instead of a set) of tuples, for ease of modelling)

int: n_Vin; 
set of int: Vin = 1..n_Vin; % source nodes on incoming edges

int: n_Ein;
array[1..n_Ein, 1..2] of int: Ein;

array[Vin] of 1..2: PARTin; 

%%%%% decision variables %%%%
array[V] of var 0..2: PART; 
var 0..(max_node_w * n_V): PART_1_size;
var 0..(max_node_w * n_V): PART_2_size;

array[1..n_Ein] of var bool: Ein_crossings; 

%%%%% constraints %%%%
% acyclic and data-dependency constraint
constraint forall (e in 1..n_E) (
    let { int : src = E [e, 1]; % local variables for readability
          int : dst = E [e, 2];
       } in

    PART [dst] = PART [src] \/
    PART [dst] = 0
  );
 
% partition sizes constraint
constraint PART_1_size = sum ( [ node_w[v] | v in V where PART [v] == 1 ] );
constraint PART_2_size = sum ( [ node_w[v] | v in V where PART [v] == 2 ] );

% inter-thread communication constraint
constraint  forall (e in 1..n_Ein) (
    let { int : src = Ein [e, 1]; % local variables for readability
          int : dst = Ein [e, 2];
       } in
    Ein_crossings [e] =
    ( PART [dst] != 0 /\
      PART [dst] != PARTin [src]
    )
  );

%%%%% objective %%%%
solve maximize 10 * min (PART_1_size, PART_2_size) - sum ( Ein_crossings );


\end{lstlisting}

\end{minipage}

\clearpage
\thispagestyle{empty}

\noindent\hspace{0.05\textwidth}
\begin{minipage}{.9\textwidth}

\begin{lstlisting} [
language=minizinc,
caption= {Parameters for the example in fig. \ref{fig:example}, to be passed as inputs to the model in listing \ref{lst:minizinc_code}},
label= {lst:minizinc_dzn_example}
]
n_V = 9;
max_node_w = 1;
node_w = [1, 1, 1, 1, 1, 1, 1, 1, 1];

n_E = 8;
E = array2d (1..n_E, 1..2, 
  [ 1, 5,
    2, 5,
    5, 7, 
    3, 6,
    4, 6,
    6, 8, 
    7, 9,
    8, 9
  ]);

n_Vin = 4; 
n_Ein = 9;

% For ease of modelling Vin are numbered 1,...,4 instead of 10,...,13 as shown in fig. 6
Ein = array2d (1..n_Ein, 1..2, 
  [ 1, 1,
    1, 4,
    1, 7,
    2, 1,
    2, 2,
    2, 8,
    3, 2, 
    3, 8,
    4, 4
  ]);

PARTin = [1, 1, 2, 2];

\end{lstlisting}

\end{minipage}

\clearpage


%



%







\end{document}